\documentclass[10pt]{iopart}

\usepackage{graphicx}
\usepackage{amssymb}
\usepackage{xcolor}

\usepackage{iopams} 
\usepackage{listings}
\usepackage{float}
\restylefloat{table}

\eqnobysec
\newcommand{\beq}{\begin{equation}}
\newcommand{\beqa}{\begin{eqnarray}}
\newcommand{\eeq}{\end{equation}}
\newcommand{\eeqa}{\end{eqnarray}}

\renewcommand{\d}{{\rm d }}

\renewcommand{\max}{{\rm max}}

\newcommand{\pro}{{\rm Prob}}

\renewcommand{\r}{{\rho}}

\newcommand{\w}{{\alpha}}

\newcommand{\fI}{f^{\rm I}}
\newcommand{\fII}{f^{\rm II}}
\newcommand{\fIII}{f^{\rm III}}
\newcommand{\FI}{F^{\rm I}}
\newcommand{\FII}{F^{\rm II}}

\newcommand{\FIII}{F^{\rm III}}
\newcommand{\LI}{\l_{\max}^{\rm I}}
\newcommand{\LII}{\l_{\max}^{\rm II}}
\newcommand{\LIII}{\l_{\max}^{\rm III}}
\newcommand{\QI}{Q^{\rm I}}
\newcommand{\QII}{Q^{\rm II}}
\newcommand{\QIII}{Q^{\rm III}}

\renewcommand{\l}{\ell}

\newcommand{\lap}[1]{\mathrel{\mathop{\cal L}\limits_{#1}^{}}}
\newcommand{\zeq}[1]{\mathrel{\mathop{=}\limits_{#1}^{}}}
\newcommand{\zapprox}[1]{\mathrel{\mathop{\approx}\limits_{#1}^{}}}
\newcommand{\erf}{\mathop{\rm erf}}

\begin{document}

\title{Statistics of the longest interval in renewal processes}
\date{\today}
\author{Claude Godr\`eche$^1$, Satya N. Majumdar$^2$ and Gr{\'e}gory Schehr$^2$}
\address{
$^1$Institut de Physique Th\'eorique, Saclay, CEA and CNRS,
91191 Gif-sur-Yvette, France}\smallskip
\address{
$^2$Laboratoire de Physique Th\'eorique et Mod\`eles Statistiques, UMR 8626, Universit\'e Paris Sud and CNRS, B\^at.~100, 91405 Orsay, France}

\begin{abstract}
We consider renewal processes where events, which can for instance be the zero crossings of a stochastic process,
occur at random epochs of time.
The intervals of time between events, 
$\tau_{1},\tau_{2},\ldots$, are independent and identically
distributed (i.i.d.) random variables with a common density $\rho(\tau)$.
Fixing the total observation time to $t$ induces a global constraint on the sum of these random intervals, which accordingly become interdependent.
Here we focus on the largest interval among such a sequence on the fixed time interval $(0,t)$. 
Depending on how the last interval is treated, we consider three different situations, indexed by $\alpha=$ I, II and III.
We investigate the distribution of the longest interval $\ell^\alpha_{\max}(t)$ and the probability $Q^\alpha(t)$ that the last interval is the longest one. 
We show that if $\rho(\tau)$ admits a well defined first moment, i.e., if it decays faster than $1/\tau^2$ for large 
$\tau$, then the full statistics of $\ell^\alpha_{\max}(t)$ is given, in the large $t$ limit, by the standard theory of extreme value statistics for i.i.d. random variables, showing in particular that the global constraint on the intervals $\tau_i$ does not play any role at large times in this case. 
However, if $\rho(\tau)$ exhibits heavy tails, $\rho(\tau)\sim\tau^{-1-\theta}$ for large $\tau$, with index 
$0 <\theta<1$ (like the zero-crossings of random walks corresponding to $\theta=1/2$), we show that the fluctuations of 
$\ell^\alpha_{\max}(t)/t$ are 
governed, in the large $t$ limit, by a stationary non-trivial universal distribution (different from a Fr\'echet law) which depends on both $\theta$ and $\alpha$, which we compute exactly. 
On the other hand, $Q^{\alpha}(t)$ is generically different from its counterpart for i.i.d. variables (both for narrow or heavy tailed distributions $\rho(\tau)$). 
In particular, in the case $0<\theta<1$, the large $t$  behaviour of $Q^\alpha(t)$ gives rise to universal non-trivial constants (depending also on both $\theta$ and $\alpha$) which we compute exactly.

\end{abstract}

\maketitle

\section{Introduction}

Renewal processes are the simplest generalizations of the Poisson process~\cite{feller,cox}.
For the latter, the time intervals between successive events (e.g., the arrival of a taxi at the airport, of telephone calls, etc.) are independent and identically distributed (i.i.d.), with a common exponential distribution.
In the case of simple renewal processes the time intervals between successive events are still i.i.d. but their common distribution is chosen arbitrary.
This distribution can be narrow, as e.g., for a Gaussian or a uniform distribution, or broad with a heavy tail~\cite{feller,dynkin}.

The simplicity of their definition explains the ubiquity and the wide range of applications of renewal processes, both in probability theory and in statistical physics.
Processes where inter-arrival times are i.i.d random variables (either exactly or as a good approximation) occur for instance in first-passage problems in Markov chains, random walks, or Brownian motion~\cite{feller,cox}, 
in the flipping of a spin for a system undergoing phase ordering~\cite{bbdg98,gl2001}, related to some problems of occupation time~\cite{lamperti}, in blinking quantum dots~\cite{blink}, in persistence properties of the diffusion equation with random initial conditions \cite{diffusion,dornicG,newman}, or in related questions~\cite{dharMaj,GDS}. 
More recently, it was shown that the renewal properties which are at the heart of the theory of 
record statistics of random walks allow to obtain a large body of exact results for these 
questions~\cite{ziff,WMS2012,MSW2012,us2,SMbook}.

In the present work we address the question of the statistics of the longest interval between successive events in renewal processes, when the process is observed between times 0 and $t$. 
Some aspects of this question were studied in the past by Lamperti~\cite{Lam61} (see also \cite{Wen64}), where the situation referred to as case I in the present work was analyzed. 
Our aim is to perform a complete and thorough re-examination of this problem, extending previous studies in several directions.
In particular we discuss the behaviour of the quantities of interest according to the nature of the distribution of intervals.
In so doing, the present work marks another step forward in the systematic 
investigation of the properties of renewal processes, in the continuation 
of the study performed in~\cite{gl2001}, where a complete study of the 
statistics of a number of observables was presented (such as $N_t$, $t_N$, 
$A_t$, $E_t$ defined below), as well as the statistics of the occupation time and related quantities, for either narrow distributions of intervals (such that all moments exist), or broad distributions with index $0<\theta<2$.

An initial study of the statistics of the longest interval between successive events in renewal processes was addressed by us in~\cite{us1}, where some exact results were announced and discussed in the context of stochastic processes in nonequilibrium systems. 
Ref.~\cite{us1} was motivated, to some extent, by a previous similar question that was raised in~\cite{ziff} for the statistics of the longest lasting record in random walks, which involves a discrete time renewal process. 
More recently, the statistics of the longest lasting record has been studied in detail for symmetric random walks in \cite{us2} as well as for random walks with a drift in \cite{MSW2012}.

The questions studied in the present work naturally belong to the wide topic of extreme value statistics, which has attracted a lot of attention during these last decades, both in mathematics and in statistical physics. 
As is well known, the statistics of extremes for i.i.d. random variables is well understood thanks to the identification, in the limit of large samples, of three distinct universality classes: Gumbel, Fr\'echet and Weibull \cite{gned,gumbel}. 
However, in many physically relevant situations, the statistics of extremes shows significant deviations from the i.i.d. case and exact results are scarce.
As exemplified here, renewal processes turn out to be a good laboratory to test, through exact analytical results, the effects (i) of non identical random variables and (ii) of correlations between them on extreme value questions.

The statistics of the largest excursion or more generally of the largest segment of stochastic processes has been studied by several authors during the last years with applications encompassing spin-glass and disordered systems \cite{DF87}, the kinetics of annihilation of charged particles \cite{FIK95} 
or charged heteropolymers \cite{KE94, EK97}. 
As discussed below (see section \ref{section:caseI}), the problems that we study here, and for which we provide exact analytical results, 
are relevant to the questions addressed in \cite{KE94, EK97}, related to the largest loops of random walks, which to a large extent were studied only numerically.

In the next section we give the precise definition of our model and of the quantities of interest studied in the present work.
We then summarize our main results and put them in perspective with their counterparts for the case of i.i.d. random variables.
The following sections are devoted to the analysis of the three cases of sequences of intervals~(\ref{def_conf}) considered in the present study. 

\section{Model and results}
\label{sec:model}

\begin{figure}
\begin{center}
\includegraphics[angle=0,width=0.9\linewidth]{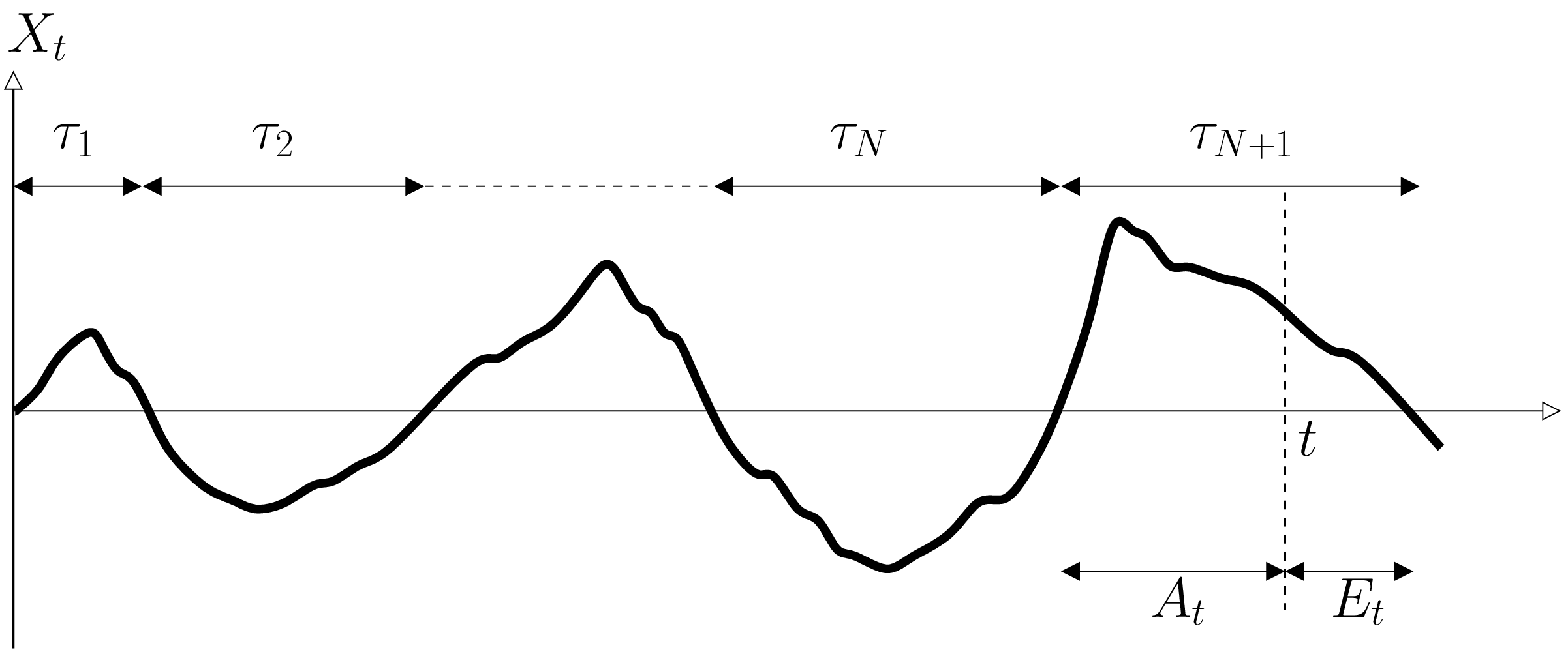}
\caption{\label{fig:renewal}
Illustration of a renewal process where the events are the zeros of the stochastic process $X_t$. 
The $\tau_i$ denote the intervals 
between the events while $A_t$ is the backward recurrence time and $E_t$ the forward recurrence time. For instance, in the case where $X_t$ is Brownian motion, the distribution of the time intervals has a power-law tail, 
$\rho(\tau) \sim \tau^{-1-\theta}$, with $\theta = 1/2$.}
\end{center}
\end{figure}

Following~\cite{gl2001}, let us consider events occurring at the random epochs of
time $t_{1},t_{2},\ldots ,$ from some time origin $t=0$. 
These events are for instance the zero crossings of some stochastic process (see figure~\ref{fig:renewal}). 
We take the origin of time on a zero crossing. 
This process is known as
a \textit{point process}. 
When the intervals of time between events, $\tau
_{1}=t_{1},\tau_{2}=t_{2}-t_{1},\ldots $, are independent and identically
distributed random variables with common density $\rho (\tau)$, the process thus
formed is a \textit{renewal process}. 
Hereafter we shall use indifferently
the denominations: events, zero crossings or renewals.

%

The probability $p_0(t)$ that no event occurred up to time $t$, given that a zero crossing has occurred at time $t=0$, or persistence probability~\cite{review}, is simply given by the
tail probability: 
\begin{equation}
p_{0}(t)=\pro(\tau_1 >t)=\int_{t}^{\infty }{\rm d}\tau\,\rho (\tau). 
\label{tail}
\end{equation}
In what follows $\rho(\tau)$ will be either a narrow distribution with all
moments finite, in which case the decay of $p_{0}(t)$, as $t\rightarrow \infty $,
is faster than any power law, or a broad distribution characterized by a
power-law fall-off with index $\theta $ (or persistence exponent~\cite{dbg,bdg}): 
\beq\label{eq:p0}
p_0(t)=\int_{t}^{\infty }{\rm d}\tau\,\rho (\tau)\approx \left( \frac{\tau_{0}}{t}%
\right) ^{\theta }\qquad (0<\theta <2), \label{def_ro}
\eeq
where $\tau_{0}$ is a microscopic time scale. 
If $\theta <1$ all moments of 
$\rho(\tau) $ are divergent, while if $1<\theta <2$, the first moment 
$\left\langle \tau\right\rangle $ is finite but higher moments are
divergent. 
In Laplace space, where $s$ is conjugate to $\tau$, for a narrow
distribution we have 
\begin{equation}
\fl \lap{\tau}\rho (\tau)=\hat{\rho}(s)=\int_0^\infty{\rm d}\tau\, \e^{-s \tau}\rho(\tau)
\zeq{ s\rightarrow 0}
1-\left\langle 
\tau\right\rangle s+\frac{1}{2}
\left\langle \tau^{2}\right\rangle s^{2}+\cdots \label{ro_narrow}
\end{equation}
The above notations for the Laplace transform will be used throughout the paper. 
For a broad distribution, (\ref{def_ro}) yields 
\begin{equation}
\hat{\rho}(s)\zapprox{s\rightarrow 0}\left\{ 
\begin{array}{ll}
1-a\,s^{\theta } & (\theta <1) \\ 
1-\left\langle \tau\right\rangle s+a\,s^{\theta } & (1<\theta <2),
\end{array}
\right. \qquad \label{ro_broad}
\end{equation}
with $a=|\Gamma (1-\theta )|\tau_{0}^{\theta }$.

Let us recall the definitions of a few natural and fundamental quantities associated to a renewal process~\cite{feller,cox,gl2001}.
Firstly, we denote by $N_t$ the number of events which occurred between $0$ and $t$, i.e., the largest $n$ such that $t_n\le t$.
The time of occurrence of the last event before $t$, that is of the 
$N_t$-th event, is therefore the sum of a random number of random variables\footnote{When no ambiguity arises, we drop the time dependence of the random
variable if the latter is itself in subscript.}
\beq\label{eq:tN}
t_{N}=\tau_{1}+\cdots +\tau_{N}. 
\eeq
The backward recurrence time $A_{t}$ (see figure~\ref{fig:renewal}) is defined as the length of time
measured backwards from $t$ to the last event before $t$, i.e., 
\beq
A_{t}=t-t_{N}.
\eeq
It is therefore the age of the current, unfinished, interval at time $t$.
Finally the forward recurrence time (also called the excess time or residual time) $E_{t}$ is the time
interval between $t$ and the next event (see figure~\ref{fig:renewal}), 
\beq
E_{t}=t_{N+1}-t. 
\eeq
We have the simple relation $A_t+E_t=t_{N+1}-t_{N}=\tau_{N+1}$.

In this work, our focus is on the longest interval between two events. 
When considering such a renewal process on a fixed time interval $(0,t)$, the last interval plays a singular role 
(see figure~\ref{fig:renewal}). 
Hence, following~\cite{us2,us1}, it is natural to 
distinguish three possible sequences of intervals (or configurations) of interest, 
\beqa\label{def_conf}
{\cal C}^{\rm I} &=& \{\tau_1, \tau_2, \ldots, \tau_{N}, A_t \} ,
\nonumber\\
{\cal C}^{\rm II} &=& \{\tau_1, \tau_2, \ldots, \tau_{N}, \tau_{N+1} \}, 
\nonumber\\
{\cal C}^{\rm III} &=& \{\tau_1, \tau_2, \ldots, \tau_{N}\},
\eeqa 
which accordingly yield three cases for the longest interval,
\beqa\label{eq:def_lmax}
\LI(t)&=&\max(\tau_1, \tau_2, \ldots, \tau_{N}, A_t),\nonumber\\
\LII(t)&=&\max(\tau_1, \tau_2, \ldots, \tau_{N}, \tau_{N+1}),\nonumber \\
\LIII(t)&=&\max(\tau_1, \tau_2, \ldots, \tau_{N}),
\eeqa
as well as for the probability that this longest interval be the last one, or probability of record breaking:
\beqa\label{eq:def_Q}
\QI(t)&=&\pro (A_t > \max(\tau_1, \tau_2, \ldots, \tau_{N})),\nonumber \\
\QII(t)&=&\pro (\tau_{N+1} > \max(\tau_1, \tau_2, \ldots, \tau_{N})),\nonumber \\
\QIII(t)&=&\pro (\tau_{N} > \max(\tau_1, \tau_2, \ldots, \tau_{N-1})).
\eeqa

As mentioned above, the study of the quantities $\ell_{\max}^\w(t)$ and $Q^\w(t)$, where $\w=$ I, II or III, have been the subject of several recent studies, which we briefly summarize. 
Refs.~\cite{ziff,MSW2012} were concerned with the study of the average value $\langle \ell^{\rm I}_{\max}(t)\rangle $.
In~\cite{us1}, we focused on 
$\langle \ell^\w_{\max}(t) \rangle$ (when it exists) for all three cases $\w=$ I, II, III, as well as on $Q^{\rm I}(t)$ (thus restricting to the case $\alpha =$ I for this quantity), for power-law distributions of intervals with index 
$0<\theta<2$. 
More recently, in~\cite{us2}, we studied the quantities $\ell^\w_{\max}(t)$ and $Q^\w(t)$, for all $\w$, in the context of record statistics of random walks. 
The associated renewal process is thus defined in discrete time with tail exponent $\theta = 1/2$ (as for the excursions of Brownian motion). 

Here, we compute the full distribution of $\ell_{\max}^\w(t)$ 
for any distribution of intervals $\rho(\tau)$ and for the three cases 
$\w=$ I, II, III.
The rest of this section is devoted to a brief presentation of our main results.

Though the random variables $\tau_1,\tau_2,\ldots$, drawn from the common distribution $\rho(\tau)$, are a priori independent and identically distributed,
the induced random variables $\tau_1,\ldots,\tau_{N},\tau_{N+1},A_t$ occurring in the sequences 
${\cal C^{\w}}$ are neither all identically distributed (except for the $N_t$ first ones, as in case III) nor independent, because fixing an observation time $t$ implies the interdependence of these intervals\footnote{A discussion of this point is given in section~5
of~\cite{gl2001}.
In~(\ref{eq:tn}) the random variables are independent, in~(\ref{eq:tN}) they are not.}. 
In particular, any of the first $N_t$ intervals in ${\cal C^{\w}}$ is obviously smaller than $t$.
While, in the first situation, fixing the number $n$ of intervals implies that the sum 
\beq\label{eq:tn}
t_n=\tau_1+\cdots+ \tau_n
\eeq 
fluctuates, in the second situation, $t$ is fixed but the number of variables, $N_t$, fluctuates. 
This is reminiscent of what occurs when changing ensembles in statistical mechanics.
We are thus naturally led to put our results in perspective with their counterparts for the case of $n$ i.i.d. random intervals $\tau_i$.

Let us recall that, if
\beq
\tau_{\max}(n)=\max(\tau_1,\ldots,\tau_n)
\eeq
is the maximum of the $n$ i.i.d. positive random variables $\tau_1,\ldots,\tau_n$,
then, according to the nature of the distribution $\rho(\tau)$ of these variables, the asymptotic distribution of this maximum, after appropriate rescaling, falls in one of three classes, namely, the Gumbel class (narrow distribution with unbounded support), the Weibull class (narrow distribution with bounded support) and the Fr\'echet class (broad distribution with a power-law tail of index $\theta>0$)~\cite{gned}.
It is also well known that, for these i.i.d. variables, the probability of record breaking 
is given by~\cite{renyi}
\beq\label{eq:renyi}
Q(n)=\pro(\tau_n>\max(\tau_1,\ldots,\tau_{n-1}))=\frac{1}{n},
\eeq
irrespectively of the distribution of these random variables.

Let us first discuss the statistics of $\l_{\max}^{\w}(t)$. 
For a narrow distribution of intervals, with finite moments,
one has~\cite{feller,cox,gl2001}
\beq\label{eq:nt}
\langle N_t\rangle\approx \frac{t}{\langle\tau\rangle},
\eeq 
One thus expects asymptotic equivalence between the statistics of $\l_{\max}^{\w}(t)$
and that of its counterpart for i.i.d. random variables, $\tau_{\max}(n)$.
This is corroborated by our results.
As shown in table~\ref{tab:exp}, for the simplest renewal process with exponential distribution of intervals, $\rho(\tau)=\e^{-\tau}$, 
the logarithmic growth of $\langle \l^\w_{\max}(t)\rangle \approx \ln t+\gamma$, where $\gamma = 0.577216 \ldots$ is the Euler constant, is similar to the one obtained for $n\sim t$ i.i.d. random variables, $\langle \tau_{\max}(n)\rangle \approx \ln n+\gamma$. 
In addition, we also show that the distribution of $(\ell^\w_{\max}(t) - \ln{t})$ is given, in the large $t$ limit, by a Gumbel distribution (\ref{eq:Gumbel}), as for i.i.d. random variables. 
This correspondence can be generalized to any narrow distribution of intervals with unbounded support.
For instance, for a Gaussian distribution, we have now $\langle \l^\w_{\max}(t)\rangle \approx (\ln t/\langle\tau\rangle)^{1/2}$, which is in line with its counterpart for i.i.d. random variables where $\langle \tau_{\max}(n)\rangle \approx (\ln n)^{1/2}$, and the properly rescaled variable asymptotically follows the Gumbel law.
Likewise, for a narrow distribution with a bounded support, such as, for instance, a uniform distribution of intervals,
we have $\langle \l^\w_{\max}(t)\rangle \approx 1-\langle\tau\rangle/t$, which is in line with its counterpart for i.i.d. random variables where $\langle \tau_{\max}(n)\rangle \approx 1-1/n$, and the rescaled maximum is exponential (which is a special case of a Weibull distribution).

For a distribution with a power-law distribution of intervals with index $\theta>1$,
(\ref{eq:nt}) still holds, hence, again, the statistics of $\l_{\max}^{\w}(t)$ and $\tau_{\max}(n)$ are expected to be asymptotically equivalent.
This is indeed the case: 
table~\ref{tab:levy<2} gives the scaling $\langle\l_{\max}^{\w}(t)\rangle\sim t^{1/\theta}$, and as shown in the text, the rescaled variable $\l_{\max}^{\w}(t)/t^{1/\theta}$ asymptotically converges to a Fr\'echet random variable $Z^F$~(\ref{eq:frechet})
of index $\theta$, for the three cases $\w=$ I, II, III. 
Hence this shows that if the distribution $\rho(\tau)$ has a well defined first moment, i.e., if $\rho(\tau)$ decays faster than $1/\tau^2$ for large $\tau$ (this includes the case where $\rho(\tau)$ vanishes identically beyond a certain value $\tau_{M}$), the limiting distribution of $\ell_{\max}(t)$, properly shifted and scaled, is the same as for i.i.d. random variables. 
This implies that neither the global constraint on the sum of the intervals $\tau_i$ nor the fact that the sequences ${\cal C}^\alpha$ (\ref{def_conf}) consists of non-identical variables (for the cases $\alpha = {\rm I}$ and $\alpha = {\rm II}$) plays a role in the statistics of $\ell_{\max}(t)$ for large $t$.

\begin{table}[h*]
\caption{Asymptotic results at large times for an exponential distribution of intervals $\rho(\tau)$, 
where $\gamma$ is the Euler constant.}
\label{tab:exp}
\begin{center}
\begin{tabular}{|c||c|c|}
\hline
$\w$&$\langle\l_{\rm max}^{\w}(t)\rangle$&$Q^{\w}(t)$\\
\hline
I&$\approx\ln t+\gamma$&$\approx1/t$\\
II&$\approx\ln t+\gamma$&$\approx \ln t/t$\\
III&$\approx\ln t+\gamma$&$\approx1/t$\\
\hline
\end{tabular}
\end{center}
\end{table}
\begin{table}[ht]
\caption{Asymptotic results at large times for a broad distribution of intervals $\rho(\tau)$ with $\theta>1$.}
\label{tab:levy<2}
\begin{center}
\begin{tabular}{|c||c|c|}
\hline
$\w$&$\langle\l_{\rm max}^{\w}(t)\rangle$&$Q^{\w}(t)$\\
\hline
I&$\sim t^{1/\theta}$&$\sim t^{1/\theta-1}$\\
II&$\sim t^{1/\theta}$&$\sim t^{1/\theta-1}$\\
III&$\sim t^{1/\theta}$&$\approx{\langle\tau\rangle}/{t}$\\
\hline
\end{tabular}
\end{center}
\end{table}

The situation is quite different, and more interesting from the point of view of the statistics of extremes, in the case where the distribution of time intervals has a power-law tail $\rho(\tau) \sim \tau^{-1 - \theta}$ with index $0<\theta<1$. 

Let us first present a heuristic argument yielding the typical  behaviour of $\ell_{\max}(t)$ as a function of time. 
For i.i.d. random variables
it was pointed out by L\'evy~\cite{levy} that the largest term of the sum outshadows the contribution of all the other terms.
This statement can be made more precise as follows.
The sum~(\ref{eq:tn}) of the $n$ i.i.d. positive random variables $\tau_1,\ldots,\tau_n$ scales as
\beq\label{eq:scale}
t_n\sim n^{1/\theta}X(\theta,1),
\eeq
denoting by $X(\theta,1)$ the one-sided stable law of index $\theta$ (and asymmetry parameter equal to $1$).
On the other hand,
\beq\label{eq:frech}
\tau_{\max}(n)\sim n^{1/\theta}Z^{F}
\eeq
where $Z^{F}$ is a random variable.
Hence,
the rescaled variable $t_n/\tau_{\max}(n)$ is expected to have a limiting distribution, denoted by $f_W$, corresponding to the random variable $W=X(\theta,1)/Z^{F}$ (where $X(\theta,1)$ and $Z^{F}$ are not independent).
This is actually the case, as shown by Darling~\cite{darling}.
The characteristic function of $W$, given in~\cite{darling} (theorem 5.1), is easily translated in Laplace space~\cite{feller} (see~(\ref{eq:darling}) below).

Let us now perform a parallel reasoning for our study, arguing as follows.
The scaling between $N_t$ and $t$ reads:
\beq\label{eq:gl}
N_t\sim t^{\theta}Y_t,
\eeq
where $Y_t$ has a limiting distribution, $Y_t\to X(\theta,1)^{-\theta}$, 
where $X(\theta,1)$ is the one-sided stable law of index $\theta$ mentioned above (see~\cite{gl2001} for a simple proof).
Using~(\ref{eq:gl}) in order to translate $n$ into $t$, and $\tau_{\max}(n)$ into $\l_{\max}(t)$, we thus infer the scaling behaviour
\beq\label{predict}
\l_{\max}(t)\sim t\, Y_t^{1/\theta} Z^{F}\sim t\frac{Z^{F}}{X(\theta,1)},
\eeq
implying the existence of the same limiting distribution for the rescaled variable $t/\l_{\max}(t)$
as for $t_n/\tau_{\max}(n)$.
Though it is difficult to predict from this heuristic reasoning whether the quantity $\l_{\max}(t)$ thus defined can be identified to one of the observables $\l_{\max}^{\w}(t)$, if any,
it nevertheless gives an argument in favour of the existence of limiting distributions for the ratios $t/\l_{\max}^{\w}(t)$.

Turning to our results, the predicted scaling~(\ref{predict}) is corroborated by table~\ref{tab:levy<1}.
Furthermore, the rescaled variables $V^{\w}_t=t/\l^{\w}_{\max}(t)$ have indeed limiting distributions, given in Laplace space respectively by~(\ref{eq:basisI}) for case I,~(\ref{eq:basisII}) for case II and~(\ref{eq:basisIII}) for case III, and depicted in figure~\ref{fig:laplace_V_I}.
Figures~\ref{fig:fV_I} (case I) and~\ref{fig:fV_III} (case III) depict these distributions in real space, while figures~\ref{fig:fR_I},~\ref{fig:fR_II},~\ref{fig:fR_III} 
depicts the distributions of their inverses (all these figures for $\theta = 1/2$).
A striking fact is that {\it the distribution of Darling, $f_W$, for i.i.d. variables, is identical to our prediction for $\fI_V$}, given in Laplace space by~(\ref{eq:basisI}).
There is thus asymptotic equivalence between $t/\l^{\rm I}_{\max}(t)$ for case I and $t_n/\tau_{\max}(n)$ for the case of i.i.d random intervals. 
The distribution of $t/\ell^{\rm I}_{\max}(t)$ was previously computed by Lamperti~\cite{Lam61}, and its connection with the distribution of Darling is pointed out in~\cite{pitman}. 
We provide here a simple alternative method, different from the one of \cite{Lam61}, to compute the distribution of $t/\ell^{\rm I}_{\max}(t)$.

\begin{table}[ht]
\caption{Asymptotic results at large times for a broad distribution of intervals $\rho(\tau)$ with $\theta=1/2$.
The constants $A$ and $B$ are respectively equal to $1/2$ and $\gamma/2+\ln (2/\pi)\approx0.4094\dots$.}
\label{tab:levy<1}
\begin{center}
\begin{tabular}{|c||c|c|}
\hline
$\w$&$\langle\l_{\rm max}^{\w}(t)\rangle/t$&$Q^{\w}(t)$\\
\hline
I&$\approx0.6265\dots$&$\approx0.6265\dots$\\
II&$\infty$&$\approx0.8001\dots$\\
III&$\approx 0.2417\dots$&$\approx(A\ln t+B)/t^{1/2}$\\
\hline
\end{tabular}
\end{center}
\end{table}

Let us close by discussing our results concerning the probability of record breaking $Q^{\alpha}(t)$.
Tables~\ref{tab:levy<2} and~\ref{tab:levy<1} show that, in general, this probability
is different from its counterpart $Q(n)$ for i.i.d. variables~(\ref{eq:renyi}).
This difference can be seen for cases I and II, except in the special case of an exponential distribution, as shown in table~\ref{tab:exp} where there is asymptotic equivalence with the i.i.d. situation, for case I, but not even for case II.
For any other narrow distribution this equivalence is lost.
In case III, $\tau_1,\ldots, \tau_{N}$ are exchangeable, hence $\QIII(t)=\langle 1/N_t\rangle$, which translates into asymptotic equivalence with $Q(n)$, not complete yet, since, as shown in table~\ref{tab:levy<1}, there is a logarithm in the numerator of $\QIII(t)$.

Let us note that for $0<\theta<1$, the asymptotic values of $Q^\w(t)$ are characterized by non-trivial universal constants for cases I and II, but not for case III (see table~\ref{tab:levy<1}).

\section{Case I}\label{section:caseI}

\subsection{Distribution of $\LI(t)$}

We first expound the general method, then discuss the results according to the nature of the distribution $\rho(\tau)$ of intervals.

We begin by determining the distribution function of 
\beq
\LI(t)=\max(\tau_1,\ldots,\tau_N,A_t),
\eeq
denoted by
\beq
\FI(t;\l)=\pro(\LI(t)\le\l).
\eeq
This function bears an explicit dependence in time $t$, which plays a fundamental role in the definition of the renewal process and in the discussion hereafter, besides its dependence in the temporal variable $\l$ associated to $\LI(t)$.
We consider the joint probability distribution of $\LI(t)$ and $N_t$,
\beqa
\FI_n(t;\l)
=\pro(\LI(t)\le\l,N_t=n)
\nonumber\\
=\int_{0}^{\l}{\rm d}\l_1\,\ldots\int_0^\l{\rm d}\l_n\,\int_0^\l{\rm d}a\,f^{{\rm I}}(t;\l_1,\ldots,\l_n,a)
\nonumber\\
=\int_{0}^{\l}{\rm d}\l_1\,\r(\l_1)\ldots\int_0^\l{\rm d}\l_n\,\r(\l_n)\int_0^\l{\rm d}a\,p_0(a)
\,\delta\left(\sum_{i=1}^n\l_i+a-t\right), 
\label{eq:F1_first}
\eeqa
where $\{\l_1,\ldots,\l_n,a\}$ is a realization of the configuration ${\cal C}^{\rm I}$,
and $f^{{\rm I}}(t;\l_1,\ldots,\l_n,a)$, the joint density of the random variables of interest, is defined in the appendix (see~(\ref{eq:fI}) and~(\ref{eq:bref})) while $p_0(a)$ is defined in~(\ref{tail}). 
In Laplace space with respect to time, (\ref{eq:F1_first}) reads (using the notations defined in (\ref{ro_narrow})):
\beq
\lap t \FI_n(t;\l)= \hat \FI_n(s;\l)=
\left(\int_0^{\l}{\rm d}\tau\r(\tau)\e^{-s\tau}\right)^n
\int_0^\l{\rm d}a\,p_0(a)\e^{-s a},
\eeq
thus
\beq\label{eq:hatFI}
\hat \FI(s;\l)=\sum_{n\ge0}\hat \FI_n(s;\l)
=\frac{\int_0^\l{\rm d}a\,p_0(a)\e^{-s a}}{1-\int_0^{\l}{\rm d}\tau\r(\tau)\e^{-s\tau}}.
\eeq
Normalization of the distribution of $\LI(t)$ can be checked on this equation by letting $\l$ go to infinity.
We denote the integrals appearing in the right side of~(\ref{eq:hatFI}) by
\beqa\label{eq:IJ}
\fl I(s;\l)=\int_0^\l{\rm d}a\,p_0(a)\e^{-s a},\qquad
J(s;\l)=\int_0^{\l}{\rm d}\tau\r(\tau)\e^{-s\tau},
\eeqa
obeying the relation
\beq\label{eq:parties}
J(s;\l)=1-p_0(\l)\e^{-s\l}-sI(s;\l),
\eeq
obtained by an integration by parts.
It follows that
\beq
\hat \FI(s;\l)=\frac{I(s;\l)}{1-J(s;\l)}
=\frac{I(s;\l)}{p_0(\l)\e^{-s\l}+sI(s;\l)},
\eeq
and finally, in Laplace space, the complementary distribution function of $\LI(t)$, namely $1-\FI(t;\l)=\pro(\ell_{\max}(t) >\ell)$, reads
\beqa\label{eq:FI}
\fl\frac{1}{s}-\hat \FI(s;\l) 
=\frac{1}{s}\,\frac{p_0(\l)\e^{-s\l}}{p_0(\l)\e^{-s\l}+sI(s;\l)}
=\frac{1}{s}\,\frac{1}{1+sI(s;\l)\e^{s\l}/p_0(\l)}.
\eeqa
From~(\ref{eq:FI}) we can easily extract the first moment of $\LI(t)$ by noting that
\beq
\langle\LI(t)\rangle=\int_{0}^{\infty}{\rm d}\l\,(1-\FI(t;\l)).
\eeq
Hence, in Laplace space, 
\beqa\label{eq:lmax}
\lap t\langle\LI(t)\rangle&=&\int_{0}^{\infty}{\rm d}\l\,\left(\frac{1}{s}-\hat \FI(s;\l)\right),
\nonumber\\
&=&\frac{1}{s}\int_{0}^{\infty}{\rm d}\l\,\frac{1}{1+sI(s;\l)\e^{s\l}/p_0(\l)}.
\label{eq:lmax2}
\eeqa

Eqs.~(\ref{eq:FI}) and~(\ref{eq:lmax}) will be analyzed below, according to the nature of the distribution of intervals $\rho(\tau)$.

\subsection{Probability of record breaking}
The second quantity of interest is the probability that the last interval in the sequence ${\cal C}^{\rm I}$ is the longest one:
\beq
\QI(t)=\pro(\LI(t)=A_t)=\pro(A_t>\max(\tau_1,\ldots,\tau_N)).
\eeq
Its computation is similar to what was done above for $\LI(t)$.
Let 
\beq
\QI(t)=\sum_{n\ge0}\QI_n(t),
\eeq
where
\beqa
\QI_n(t)=\pro(A_t>\max(\tau_1,\ldots,\tau_N),N_t=n),
\nonumber\\
\fl =\int_0^\infty{\rm d}a\,p_0(a)
\int_{0}^{a}{\rm d}\l_1\,\r(\l_1)\ldots\int_0^a{\rm d}\l_n\,\r(\l_n)
\,\delta\left(\sum_{i=1}^n\l_i+a-t\right).
\eeqa
In Laplace space we have
\beq
\lap t \QI_n(t)=\hat \QI_n(s)=\int_0^\infty{\rm d}a\,p_0(a)\e^{-s a}\left(\int_0^{a}{\rm d}\tau\r(\tau)\e^{-s\tau}\right)^n,
\eeq
and therefore, summing on $n$,
\beqa
\fl \hat \QI(s)=\int_0^\infty{\rm d}a\, \frac{p_0(a)\e^{-s a}}{1-\int_0^{a}{\rm d}\tau\r(\tau)\e^{-s\tau}}
=\int_0^\infty{\rm d}a\, \frac{p_0(a)\e^{-s a}}{p_0(a)\e^{-sa}+sI(s;a)}.
\eeqa
We note from~(\ref{eq:FI}) and (\ref{eq:lmax}) that
\beq
\hat \QI(s)
=s \lap t\langle\LI(t)\rangle,
\eeq
yielding
\beq\label{identity}
\QI(t)=\frac{\d}{{\rm d}t}\langle\LI(t)\rangle,
\eeq
since $\langle\LI(0)\rangle=0$.

This relationship, which only holds for case I, is actually more general, and applies for any distribution of intervals, i.e., non necessarily generated by a renewal process.
Indeed, if $t$ increases by ${\rm d}t$, then either $\LI(t)$ increases by ${\rm d}t$, if the last interval, namely $A_t$, is the longest one, which happens with probability $\QI(t)$; 
or stays the same, with probability $1-\QI(t)$.
Taking the average leads to~(\ref{identity})~\cite{us1}.

\subsection{Discussion}
We now discuss the above results according to the nature of the distribution $\rho(\tau)$ of the intervals.

\smallskip\noindent{\it (i) Exponential distribution of intervals}

Let us first consider the special case of an exponential distribution of intervals, $\r(\tau)=\e^{-\tau}$ ($\tau>0$).
This corresponds to the simplest renewal process, where the events form a Poisson process. 
The starting point of our analysis is~(\ref{eq:FI}) where the functions $I(s;\l)$ and $J(s;\l)$ are given in~(\ref{eq:IJ}), (\ref{eq:parties}). 
In this case, $p_0(a)=\e^{-a}$, hence $J(s;\l)=I(s;\l)$, and we obtain
\beq\label{eq:F_I_exp}
\frac{1}{s}-\hat \FI(s;\l)=\frac{1+s}{s}\frac{1}{1+s\,\e^{(1+s)\l}}.
\eeq
Thus from~(\ref{eq:lmax}), one has
\beq\label{eq:expI}
\lap t \langle\LI(t)\rangle=\frac{1+s}{s}\int_0^\infty {\rm d}\l\,\frac{1}{1+s\,\e^{(1+s)\l}}
=\frac{1	}{s}\ln\left(1+\frac{1}{s}\right).
\eeq
By inverting the Laplace transform, we obtain
\beq\label{eq:lmax_I_exp}
\langle\LI(t)\rangle={\rm E}(t)=
\int_0^t{\rm d}u\frac{1-\e^{-u}}{u}=\sum_{k\ge1}\frac{(-1)^{k-1}}{k}\frac{t^k}{k!},
\eeq
where ${\rm E}(t)$ is defined by the second equality.
At large times we have ${\rm E}(t)\approx \ln t+\gamma$, where $\gamma$ is the Euler constant.

The asymptotic distribution of $\LI(t)$ is actually related to the Gumbel distribution.
Indeed from~(\ref{eq:F_I_exp}), we have, for $s$ small and $\l$ large,
\beq
\hat \FI(s;\l)\approx \frac{1}{s+\e^{-\l}},
\eeq
hence
\beq\label{eq:Gumbel}
\FI(t;\l)\approx\e^{-\e^{-\l}t}=\e^{-\e^{-(\l-\ln t)}}.
\eeq
In other words, we have, asymptotically,
\beq\label{eq:lI}
\LI(t)\approx \ln t+Z^{G},
\eeq
where $Z^{G}$ follows the standard Gumbel distribution, with $\langle Z^{G}\rangle=\gamma$, in agreement with what was found above.

From~(\ref{identity}) and~(\ref{eq:lmax_I_exp}) we obtain 
\beq\label{eq:QI}
\QI(t)=\frac{1-\e^{-t}}{t}\approx\frac{1}{t}.
\eeq
The right sides of~(\ref{eq:lI}) and (\ref{eq:QI}) are similar to the forms corresponding to $n$ i.i.d. exponential random variables, if one replaces $t$ by $n$.
This stems from the fact that for an exponential distribution of intervals, $A_t$ has the same distribution as the other intervals, and furthermore the number of intervals $N_t\approx t$, at large times ($\langle\tau\rangle=1$ here).
However, $N_t$ is distributed, and therefore the intervals are not strictly i.i.d. random variables.
 In summary, in the exponential case, $\LI(t)$ behaves asymptotically as the largest of $N_t\sim t$ i.i.d random variables, and $\QI(t)$ scales as $1/\langle N_t\rangle\approx1/t$.

One can check that, for a Gaussian distribution, $\langle\LI(t)\rangle\sim (\ln t)^{1/2}$, a result which is asymptotically equivalent to its counterpart for i.i.d. Gaussian variables.
Instead, $\QI(t)\sim 1/(t\sqrt{\ln t})$, obtained by derivation of 
$\langle\LI(t)\rangle$ with respect to $t$,
is not asymptotically equivalent to $1/n$.
For a uniform distribution of interval, one would find an exponential distribution for the rescaled maximum:
\beq
\l_{\max}(t)\approx 1-\frac{\langle\tau\rangle}{t}Z^{Exp},
\eeq
where $Z^{Exp}$ is exponentially distributed, in line with what is expected from the knowledge of the i.i.d. case.
And again $\QI(t)$ is not equivalent to $Q(n)$.
The same treatment can be given for any distribution in the Weibull class.
%
\begin{figure}
\begin{center}
\includegraphics[angle=0,width=0.9\linewidth]{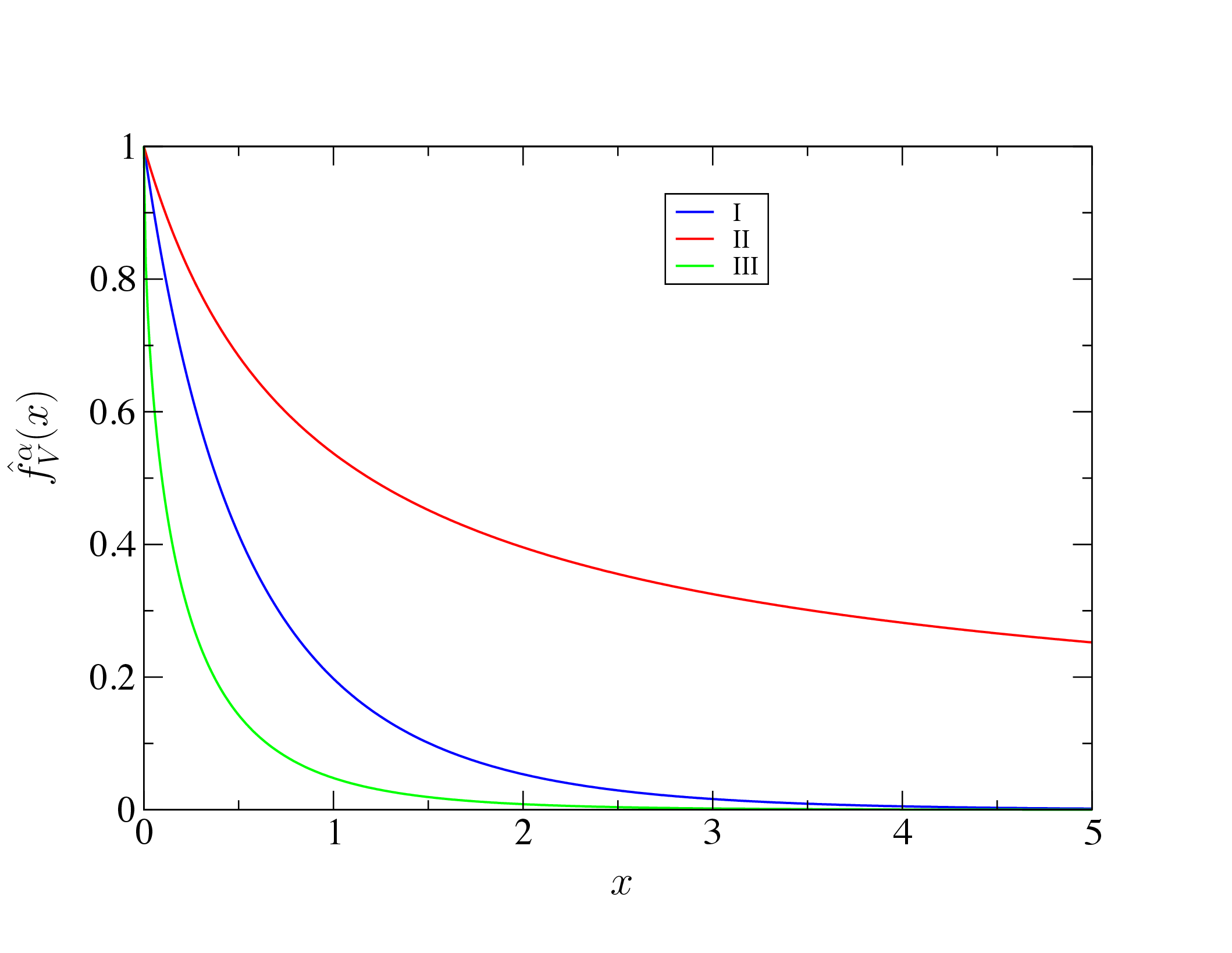}
\caption{\label{fig:laplace_V_I}
Laplace transforms $\hat f^\w_V(x)=\langle\e^{-xV}\rangle$ of the scaling functions $f^\w_V(v)$ for the three cases $\w=$ I, II, III and $\theta=1/2$ (see~(\ref{eq:basisI}), (\ref{eq:basisII}) and~(\ref{eq:basisIII})).
From top to bottom: II, I, III.
The properties of these functions are discussed in the text.
}
\end{center}
\end{figure}
\begin{figure}
\begin{center}
\includegraphics[angle=0,width=0.9\linewidth]{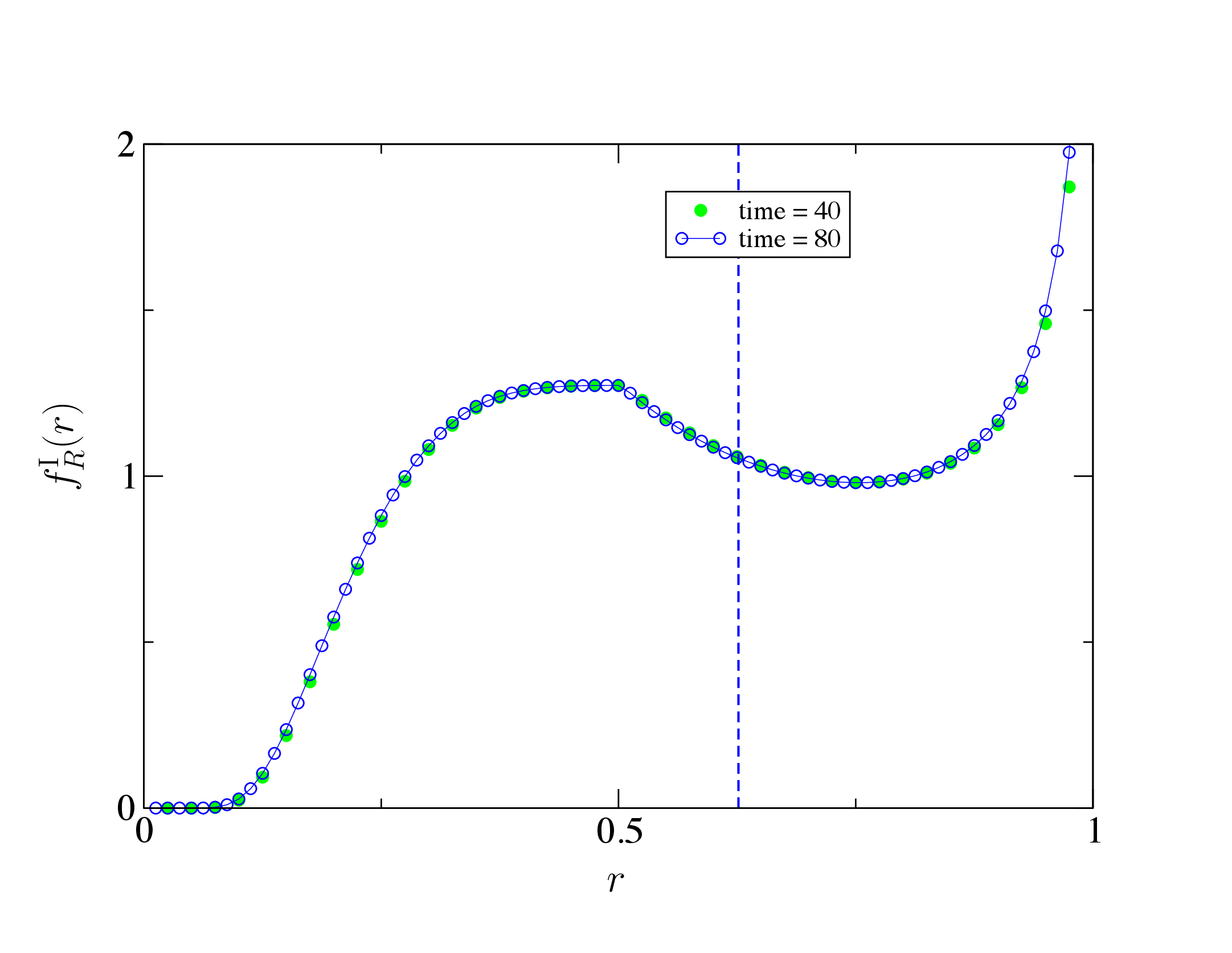}
\caption{\label{fig:fR_I}
Limiting density $\fI_R(r)$ of the scaling variable $R_t=\LI(t)/t$ at large times, for $\theta=1/2$.
\textcolor{black}{This was obtained from the analytical expression of the generating function of $\LI(t)$ for a random walk, after $40$ steps (green full circles) and $80$ steps (blue empty circles), while the line is a guide to the eyes, connecting the blue circles. 
The good collapse of the data 
demonstrates that the scaling is already very good for such values of the number of steps (see text).}
The vertical dotted line indicates the value of the first moment $\langle R\rangle\approx 0.626\,508\dots$ (see~(\ref{eq:pitman}), (\ref{eq:resultI})).
}
\end{center}
\end{figure}

\begin{figure}
\begin{center}
\includegraphics[angle=0,width=0.9\linewidth]{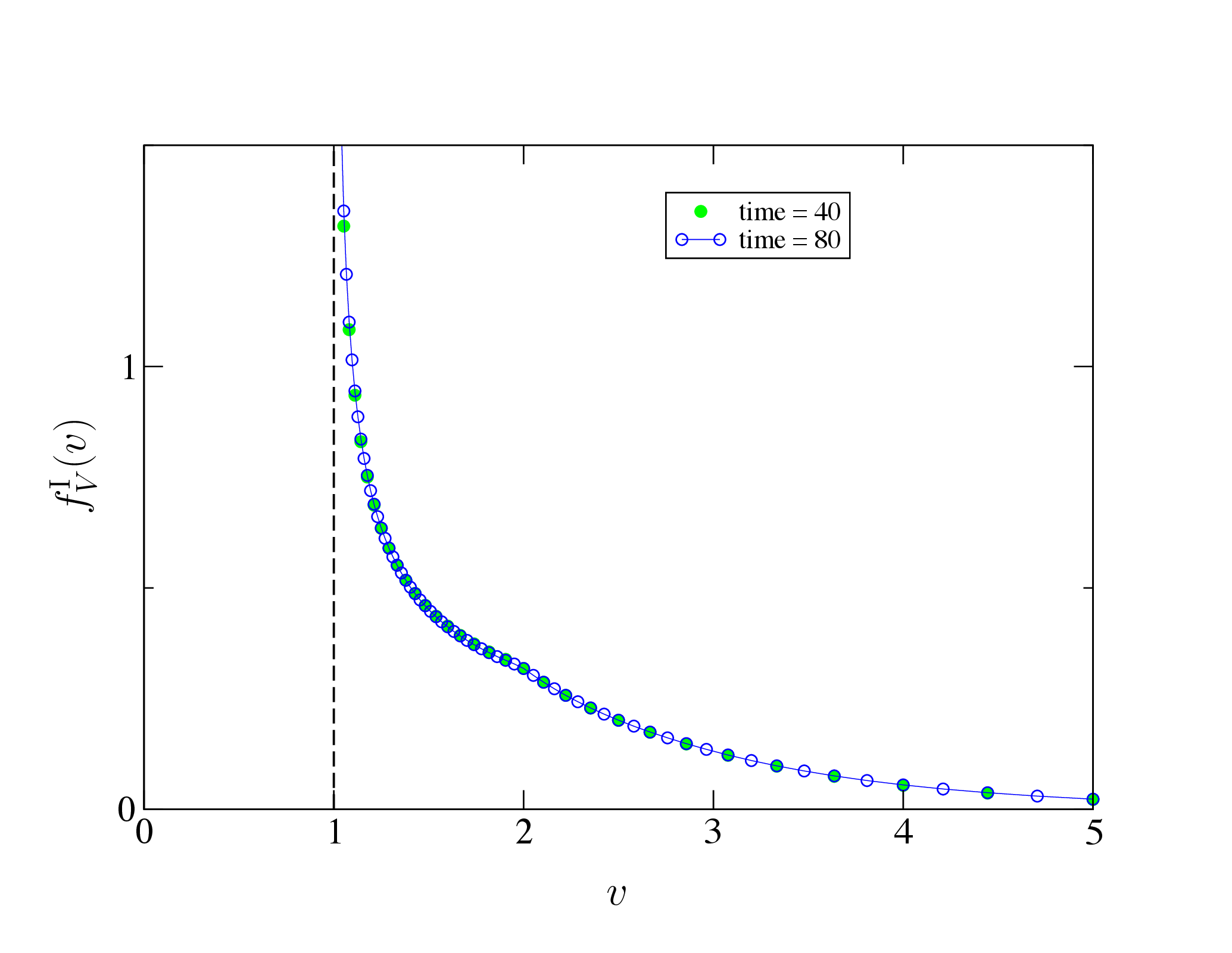}
\caption{\label{fig:fV_I}
Limiting probability density $\fI_V(v)$, where $V_t=t/\LI(t)$ as $t\to\infty$ for $\theta=1/2$, obtained from the data of figure~\ref{fig:fR_I} (using $V_t = 1/R_t$).
The tail of this function is exponential, with decay rate approximately equal to $0.854$ (see text).}
\end{center}
\end{figure}

%
\smallskip\noindent{\it (ii) Broad distribution of intervals with index $0<\theta<1$} 

\vskip 0.2cm

We now consider intervals with broad distribution $\rho(\tau)\sim 
\tau^{-1-\theta}$ for large $\tau$ (see Eq.~\ref{eq:p0}). The starting 
point of our analysis is again~(\ref{eq:FI}) 
with $I(s;\l)$ and $J(s;\l)$ given in~(\ref{eq:IJ}), (\ref{eq:parties}). 
We find that in the limit $s \to 0$, $\ell \to \infty$, keeping $x = s \l$ fixed, $I(s;\l)$ and $J(s;\l)$ take the scaling forms
\beqa\label{eq:scaling_IJ_1}
I(s;\l)\approx\tau_0^\theta s^{\theta-1} \int_0^{s \l}{\rm d}u\,u^{-\theta}\e^{-u},
\nonumber\\
1-J(s;\l)\approx
\tau_0^\theta s^{\theta} \left((s \l)^{-\theta}\e^{-s \l}+\int_0^{s \l}{\rm d}u\,u^{-\theta}\e^{-u}\right).
\eeqa
Hence by injecting these scaling forms (\ref{eq:scaling_IJ_1}) into~(\ref{eq:FI}) one finds 
\begin{eqnarray}\label{eq:basisI0}
\frac{1}{s}-\hat \FI(s;\l)\approx \frac{1}{s} \hat \fI_V(s \l),
\end{eqnarray}
with
\begin{eqnarray}\label{eq:basisI}
\hat \fI_V(x) = \frac{1}{1+x^\theta\e^x\int_0^x{\rm d}u\,u^{-\theta}\e^{-u}} ,
\end{eqnarray}
where, as we show below, the scaling function $\hat \fI_V(x)$ has a natural probabilistic interpretation.
This function can also be written as
\begin{eqnarray}\label{hypergeo}
\hat \fI_V(x) = \frac{1}{_1 F_1(1,1-\theta,x)} ,
\end{eqnarray} 
where $_1 F_1(1,1-\theta,x)$ is a confluent hypergeometric function, simply related to 
the incomplete gamma function
\beq
\Gamma(a,x)=\int_x^\infty{\rm d}u\,u^{a-1}\e^{-u}.
\eeq
For $\theta=1/2$,~(\ref{eq:basisI}) reduces to
\beq\label{eq:fhatV}
\hat \fI_V(x)=\frac{1}{1+\sqrt{\pi x}\,\e^x\erf \sqrt{x}},
\eeq
which is plotted in figure~\ref{fig:laplace_V_I}.

From~(\ref{eq:basisI}) we deduce the following.
First, using~(\ref{eq:lmax}), analyzed in the small $s$ limit (corresponding to large time), 
we have
\beq
\lap t \langle\LI(t)\rangle\approx\frac{1}{s^2}\int_0^\infty{\rm d}x\,
\frac{1}{1+x^\theta\e^x\int_0^x{\rm d}u\,u^{-\theta}\e^{-u}},
\eeq
and therefore
\beqa\label{eq:qinfty}
\fl\lim_{t\to\infty}\frac{1}{t} \langle\LI(t)\rangle
=\QI_{\infty}
=\int_0^\infty{\rm d}x\,\hat f_V(x)
=\int_0^\infty{\rm d}x\,
\frac{1}{1+x^\theta\e^x\int_0^x{\rm d}u\,u^{-\theta}\e^{-u}}.
\eeqa
In particular, for $\theta=1/2$, (\ref{eq:qinfty}) yields: 
\beq\label{eq:pitman}
\QI_\infty=\int_0^\infty{\rm d}x\,\frac{1}{1+\sqrt{\pi x}\,\e^x\erf \sqrt{x}}=0.626\,508\ldots,
\eeq
recovering a result of Pitman and Yor~\cite{pitman} (see also~\cite{finch})\footnote{Note that (\ref{eq:qinfty}) was used in~\cite{GRS} to demonstrate that $\ell_{\max}^{\rm I}(t)$ for fractional Brownian motion of Hurst index $H$ (and hence $\theta = 1-H$ \cite{review}) carries the signature of non-Markovian effects.}.

Although we naturally focused above on the variable $\l^{\rm I}_{\max}(t)/t$, it turns out that the probability density function of the random variable $t/\l^{\rm I}_{\max}(t)$ is simpler to study. 
Let us denote those two random variables by $R_t$ and $V_t = 1/R_t$:
\beq
\quad R_t=\frac{\LI(t)}{t},\quad V_t=\frac{t}{\LI(t)}.
\eeq
Then, as shown below,~(\ref{eq:basisI}) expresses the fact that in the asymptotic late-time regime the random variables
\beq\label{eq:RVasymp}
R=\lim_{t\to\infty}R_t,\quad V=\lim_{t\to\infty}V_t,
\eeq
have time-independent limiting distributions, denoted by $f^{\rm I}_R(r)$ and $f^{\rm I}_V(v)$, 
simply related by
\begin{eqnarray}\label{eq:rel_fR_fVI}
f^{\rm I}_R(r) = \frac{1}{r^2} f^{\rm I}_V\left(\frac{1}{r}\right) .
\end{eqnarray}
It turns out that the scaling function $\hat \fI_V(x)$
in~(\ref{eq:basisI}) is precisely the Laplace transform of $f^{\rm I}_V(v)$ with respect to $v$, as suggested by the notation. 
This can be simply demonstrated as follows.
At any finite time,
\beq\label{eq:finitetime}
1-\FI(t;\l)=\pro(\LI(t)>\l)=\pro(V_t<v\equiv t/\l).
\eeq
In the large time $t$ limit when $V_t$ assumes a stationary distribution $f^{\rm I}_V(v)$ one has
\begin{eqnarray}\label{eq:cumul_V}
\pro(V_t<v\equiv t/\l) \approx \int_0^{t/\l}{\rm d}v\, f^{\rm I}_V(v) . 
\end{eqnarray} 
Hence, inserting this expression into~(\ref{eq:finitetime}) and performing a Laplace transform of both sides of the latter with respect to $t$, one obtains
\beq\label{eq:cumul_V2}
\hspace*{-0.8cm}\frac{1}{s}-\hat\FI(s;\l) \approx \int_0^\infty{\rm d}t\, \e^{-s t} \int_0^{t/\l}{\rm d}v\, f_V^{\rm I}(v) 
= \frac{1}{s} \int_0^\infty{\rm d}v\, f^{\rm I}_V(v)\e^{-(s \l) v},
\eeq
where we used an integration by parts to obtain the last equality. 
Note that this result is actually valid only for small $s$ (since we used 
(\ref{eq:cumul_V}) which itself is valid only for 
large $t$).
By comparing~(\ref{eq:basisI0}) and~(\ref{eq:cumul_V2}) we finally obtain 
\begin{eqnarray}\label{eq:fvlap}
\hat \fI_V(x) = \langle\e^{-x V}\rangle=
\int_0^\infty{\rm d}v\, \e^{-x v} f^{\rm I}_V(v) .
\end{eqnarray}
Hence~(\ref{eq:basisI}) yields back the result of Lamperti \cite{Lam61}.

Using (\ref{eq:fvlap}) we can recast~(\ref{eq:qinfty}) as
\beq\label{eq:resultI}
\lim_{t\to\infty}\frac{1}{t} \langle\LI(t)\rangle=\QI_\infty
=\int_0^\infty{\rm d}x\,\hat f^{\rm I}_V(x)
=\left\langle\frac{1}{V}\right\rangle=\langle R\rangle.
\eeq
More generally, higher moments of $R$ can be obtained as
\beq
\langle R^p\rangle=\frac{1}{\Gamma(p)}\int_0^\infty{\rm d}x\,x^{p-1}\hat f^{\rm I}_V(x).
\eeq
For instance, if $\theta=1/2$, $\langle R^2\rangle=0.4565\dots$, $\langle R^3\rangle=0.3653\dots$,
which have no explicit expressions.
On the other hand the first few moments of $V$ have simple rational expressions (see also \cite{Lam61}):
\beq
\langle V\rangle=\frac{1}{1-\theta},\quad \langle V^2\rangle=\frac{2}{(1-\theta)^2(2-\theta)},\ldots . \label{eq:moments_V_I}
\eeq

As discussed in section~\ref{sec:model}, for i.i.d. random variables the ratio of the sum to the maximal term has a limiting distribution, $f_W(w)$, as found in~\cite{darling}, which is identical to $\fI_{V}(v)$.
The expression given in~\cite{feller,darling} reads
\beq\label{eq:darling}
\hat f_W(x)=\frac{\e^{-x}}{1-\theta\int_0^1{\rm d}u\,u^{-1-\theta}(\e^{-u x}-1)},
\eeq
where $W=\lim_{t\to\infty}t_n/\tau_{\max}(n)$ (see~(\ref{eq:scale}), (\ref{eq:frech})) and $x$ is the Laplace variable conjugate to $w$.
It is easily seen that $\hat f_W(x)$ given by~(\ref{eq:darling}) and $\hat\fI_V(x)$ given by~(\ref{eq:basisI}) are identical.

Let us now analyze the generic features of the distributions $\fI_R(r)$ and $\fI_V(v)$.
As in other similar problems~\cite{Lam61,Wen64,DF87,FIK95,KE94,EK97,glrecord}, one expects that $\fI_V(v)$ has a different expression on each interval $[k,k+1]$, with $k=1,2,\dots$, with a singularity at each integer $v = 2,3, \dots$ \cite{Lam61}, implying that $\fI_R(r)$ has singularities at $r = 1/2, 1/3, \dots$. 
In particular, for $1<v<2$, one has~\cite{Lam61}
\beq\label{eq:fv_smallv}
f^{\rm I}_V(v) = \frac{\sin{\pi \theta}}{\pi} (v-1)^{\theta-1} , \; 1 < v < 2 ,
\eeq
as can be seen by inspection of the large $x$ behaviour of~(\ref{eq:basisI}),
$\hat\fI_V(x)\approx x^{-\theta}\e^{-x}/\Gamma(1-\theta)$.

On the other hand, the large $v$  behaviour of $f^{\rm I}_V(v)$ is given by the analytic structure of its Laplace transform 
$\hat f^{\rm I}_V(x)$~(\ref{eq:basisI}), (\ref{hypergeo}) in the complex $x$--plane. 
The latter is a meromorphic function, with simple poles located at the zeros $x=s_k$ of the hypergeometric function 
$_1 F_1(1,1-\theta,x)$ (see e.g., \cite{Wen64}). 
These zeros are
such that $s_{\pm k} = - \alpha_k \pm i \beta_k$ (with $\alpha_k$ and $\beta_k$ real) with a negative real part ($\alpha_k > 0$ for all $k$) and $0<\alpha_0<\alpha_1< \alpha_2< \dots$. 
Furthermore, $s_0 = - \alpha_0$ is the only real zero (i.e., $\beta_0 = 0$). 
The residues of $\hat f^{\rm I}_V(x)$ at the poles $x=s_k$ are given by $-s_k/\theta$,
from which we obtain the large $v$  behaviour of $\fI_{V}(v)$ as
\begin{eqnarray}\label{eq:fv_largev}
\fl f^{\rm I}_V(v) \approx \frac{\alpha_0}{\theta} \e^{-\alpha_0 \, v} + \frac{2}{\theta} \e^{-\alpha_1 v} \left( \alpha_1 \cos{(\beta_1 v)} + \beta_1 \sin{(\beta_1 v)} \right) + {\cal O} (\e^{-\alpha_2 v}) .
\end{eqnarray}
Thus the first subleading contribution to the large $v$  behaviour contains an oscillating part. 
For $\theta = 1/2$, a numerical estimation of the roots $s_0$ and $s_1$ yields $\alpha_0 = 0.854032\ldots$, $\alpha _1 = 4.24892\ldots$ and $\beta_1 = 6.38312 \ldots$ (see also \cite{Wen64}). 
From~(\ref{eq:fv_smallv}) and (\ref{eq:fv_largev}), together with~(\ref{eq:rel_fR_fVI}), we finally obtain the asymptotic  behaviours of $\fI_R(r)$ as
\begin{eqnarray}\label{eq:asympt_f_RI}
f^{\rm I}_R(r) \approx
\left\{
 \begin{array}{lll}
 ({\alpha_0}/\theta) \, r^{-2} \e^{-{\alpha_0}/{r}} , & \; r \to 0 \\
 & \\ 
 (\sin(\pi\theta)/\pi)(1-r)^{\theta-1} , & \; r \to 1 .
 \end{array}
\right.
\end{eqnarray}

In order to have a graphical representation of the densities $\fI_R(r)$ or $\fI_V(v)$ one could either perform the numerical inversion of the Laplace transform of $\hat\fI_V(v)$, given by~(\ref{eq:basisI}), or perform a simulation of the process yielding histograms of these densities.
Restricting to the case where $\theta=1/2$, a third method consists in using the expression of the distribution function of $\LI(t)$, given as an explicit series for the case of the longest lasting record for a random walk~\cite{us2}.
This case is the discrete counterpart (where $t$ is a discrete variable) of the continuous renewal process studied in the present work. 
In~\cite{us2} (see~(28) therein) we obtained an explicit expression for the generating function $\tilde F^{\rm I}(z;\ell) = \sum_{t\geq 0} F^{\rm I}(t;\ell) z^t$. 
Hence the value of $F^{\rm I}(t;\ell)$ can be read off from the expansion of $\tilde F^{\rm I}(z;\ell)$ close to $z=0$.
This procedures makes sense as~(\ref{eq:basisI}) demonstrates that there is universality of the result in the scaling regime since the scale $\tau_0$ disappears in the expression of the densities in Laplace space.
Figures~\ref{fig:fR_I} and~\ref{fig:fV_I} depict the densities $\fI_R(r)$ and $\fI_V(v)$ obtained that way (the same method is used for cases II and III).
\textcolor{black}{The discontinuities of the first derivatives at $r=1/2$ for $\fI_R(r)$ and $v=2$ for $\fI_V(v)$ are noticeable.}

Let us finally mention that a quantity similar to (albeit different from) $\ell_{\max}^{\rm I}(t)$ was studied in~\cite{KE94,EK97} in the context of charged polymers. 
In these works, the authors studied the ``loops'' of the random walks, i.e., segments 
whose beginning and ending positions coincide,
focussing on the largest of such segments irrespectively of their starting points.
Translated into the language of, e.g., random walks, the renewal process considered in the present work corresponds to paths that start at the origin but do not necessarily end there. 
Despite this simplification, our distribution $f^{\rm I}_R(r)$ shares several features with the distribution of the largest loop studied, mainly numerically, in~\cite{KE94,EK97}: it exhibits a non-analytic  behaviour for $r=1/2, 1/3, \ldots$, an essential singularity when $r \to 0$ and a square root divergence when $r \to 1$ (see~(\ref{eq:asympt_f_RI})).

\smallskip\noindent{\it (iii) Broad distribution with index $1<\theta<2$}
\label{sec:lev_I_tet>1}

We restart from~(\ref{eq:FI})
\beq
\frac{1}{s}-\hat \FI(s;\l)
=\frac{1}{s}\,\frac{1}{1+sI(s;\l)\e^{s\l}/p_0(\l)},
\eeq
that we want to analyze in the late-time regime ($s\to0$), where $\LI(t)$ and $\l$ are large.
We have $p_0(\l)\approx (\tau_0/\l)^\theta$.
In order to avoid the divergence of $I(s;\l)$ at the lower bound, we write
\beq
I(s;\l)=\hat p_0(s)-\int_\l^\infty{\rm d}u\,p_0(u)\e^{-s u}.
\eeq
We have
\beq\label{eq:anal1}
\hat p_0(s)=\frac{1-\hat\rho(s)}{s}\approx\langle\tau\rangle-as^{\theta-1},
\eeq
and
\beq
\int_\l^\infty{\rm d}u\,p_0(u)\e^{-s u}\approx\tau_0^\theta\int_\l^\infty
{\rm d}u\,u^{-\theta}\e^{-su}
=\tau_0^\theta s^{\theta-1}\Gamma(1-\theta,s\l).
\eeq
So,
\beq\label{eq:anal2}
\frac{sI(s;\l)\e^{s\l}}{p_0(\l)}\approx \frac{\e^{s\l}s\l^\theta}{\tau_0^\theta}
(\langle\tau\rangle-s^{\theta-1}(a+\tau_0^\theta\Gamma(1-\theta,s\l)))
\eeq
that we have to further analyze.
\begin{enumerate}
\item The second term in the parentheses $s^{\theta-1}(a+\tau_0^\theta\Gamma(1-\theta,s\l))$ is subleading compared to the first one, $\langle\tau\rangle$.
\item One has to compare the two products $s\l$ and $s\l^\theta$.
It is clear that $\l\sim s^{-1/\theta}$ because then $s\l\sim s^{1-1/\theta}$ is small, while
$\l\sim s^{-1}$ yields $s\l^\theta\sim s^{1-\theta}$ which is large.

\end{enumerate}\medskip

We thus have $\e^{s\l}\approx1$. 
We are left with
\beq\label{eq:FI>1}
\frac{1}{s}-\hat \FI(s;\l)
\approx \frac{1}{s}\frac{1}{1+s\langle\tau\rangle(\l/\tau_0)^\theta},
\eeq
or
\beq
\hat \FI(s;\l)\approx\frac{1}{s+(\l/\tau_0)^{-\theta}/\langle\tau\rangle},
\eeq
yielding finally the distribution function of $\LI(t)$,
\beq
\FI(t;\l)\approx\e^{-t/\langle\tau\rangle(\l/\tau_0)^{-\theta}}.
\eeq
Setting
\beq
\LI(t)=\tau_0\left(\frac{t}{\langle\tau\rangle}\right)^{1/\theta}Z_t,
\eeq
we have, as $t\to\infty$, $Z_t\to Z^{F}$, with limiting distribution 
\beq
\pro(Z^{F}<x)=\e^{-1/x^\theta} \label{eq:frechet}
\eeq
which is the Fr\'echet law.
Therefore
\beq\label{LI>1}
\langle\LI(t)\rangle\approx\tau_0\left(\frac{t}{\langle\tau\rangle}\right)^{1/\theta}\underbrace{\langle Z^{F}\rangle}_{\Gamma(1-1/\theta)}.
\eeq
\begin{figure}
\begin{center}
\includegraphics[angle=0,width=0.9\linewidth]{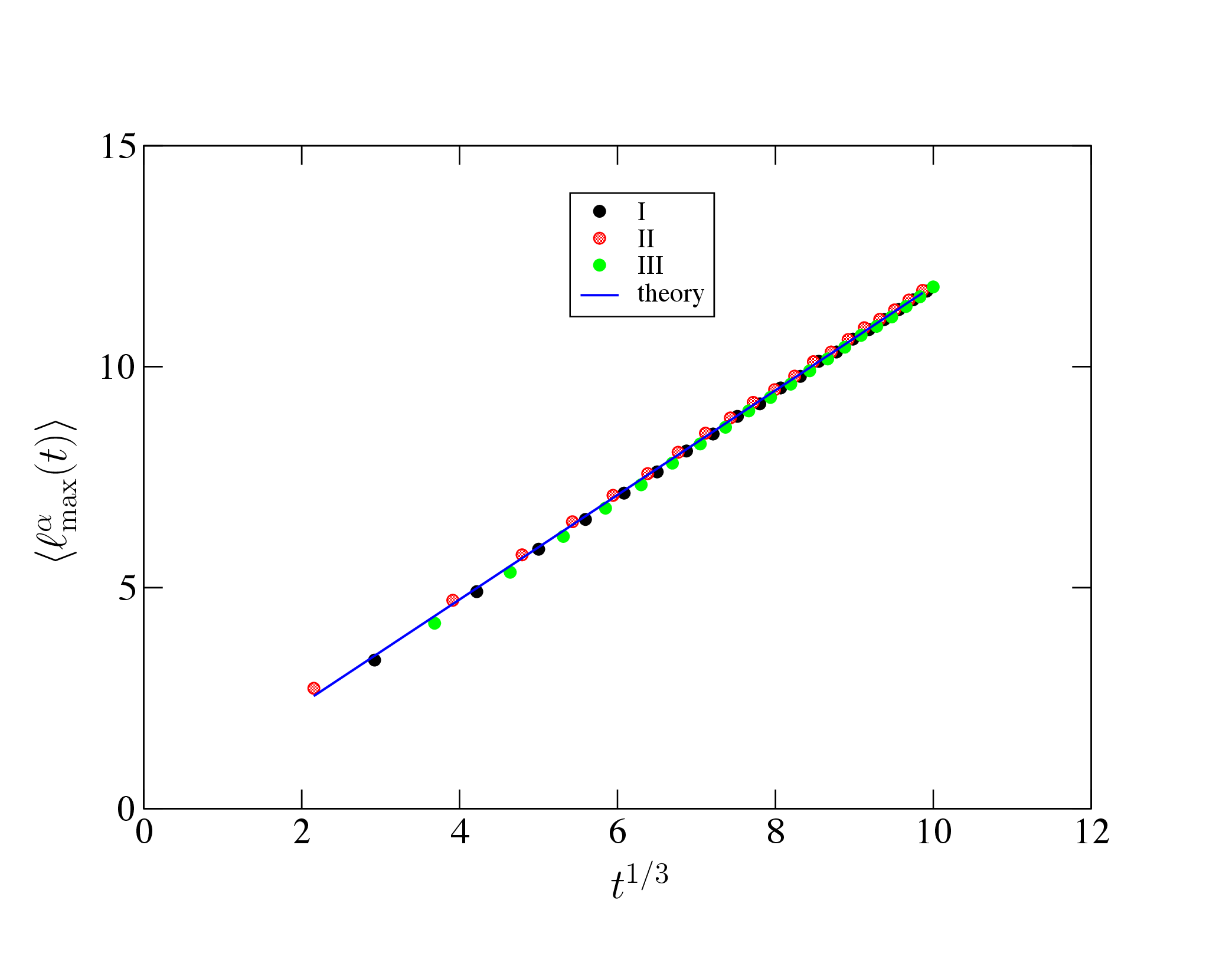}
\caption{\label{fig:lmax_I_II_III}
\textcolor{black}{Simulations of $\langle\l_{\max}^{\w}(t)\rangle$ for the cases $\alpha =$ I (black circles), $\alpha =$ II (red circles) and $\alpha=$ III (green circles) compared to the analytical prediction~(\ref{LI>1}) with $\theta=3$ (solid line).}}
\end{center}
\end{figure}
In other words, $\LI(t)$, the maximum of $\tau_1, \tau_2, \ldots, \tau_{N}, A_t$
has asymptotically the same distribution as $n$ i.i.d. random variables with common density $\rho(\tau)$.
As seen later, the same result also holds for cases II and III.
Furthermore, the analysis done above still holds for any $\theta>1$ because the existence of regular terms of higher order in~(\ref{eq:anal1}) does not affect the analysis of~(\ref{eq:anal2}).
Figure~\ref{fig:lmax_I_II_III} depicts the analytical prediction~(\ref{LI>1}) for $\theta=3$, and also compares it to simulations for all three cases.

Finally, taking the temporal derivative of~(\ref{LI>1}), we obtain the probability of record breaking
\beq\label{Q_I_levy2}
\QI(t)\approx\frac{\Gamma(1-1/\theta)\tau_0}{\theta\langle\tau\rangle}\left(\frac{t}{\langle\tau\rangle}\right)^{-(1-1/\theta)}.
\eeq
This probability recast in terms of $n\sim t/\langle\tau\rangle$
reads $\QI(t)\sim 1/n^{1-1/\theta}$, which differs from the i.i.d. result $Q(n)=1/n$.
The enhancement factor $n^{1/\theta}$ can be traced back to the role of the last interval $A_t$, as a simple calculation demonstrates.
Indeed, considering the random intervals $\tau_1, \tau_2, \ldots, \tau_{N}, A_t$ as independent, with $p_0(a)/\langle\tau\rangle$ for the density of $A_t$ (see~\cite{gl2001}), yields indeed~(\ref{Q_I_levy2}).

To close, let us underline that, when $0<\theta<1$, the results found above are universal i.e., only depend on the tail of the distribution $\rho(\tau)$, which is corroborated by the fact that the scale $\tau_0$ disappears of the expressions. 
This universality does not hold for narrow distributions or for broad distributions with $\theta>1$.

\begin{figure}
\begin{center}
\includegraphics[angle=0,width=0.9\linewidth]{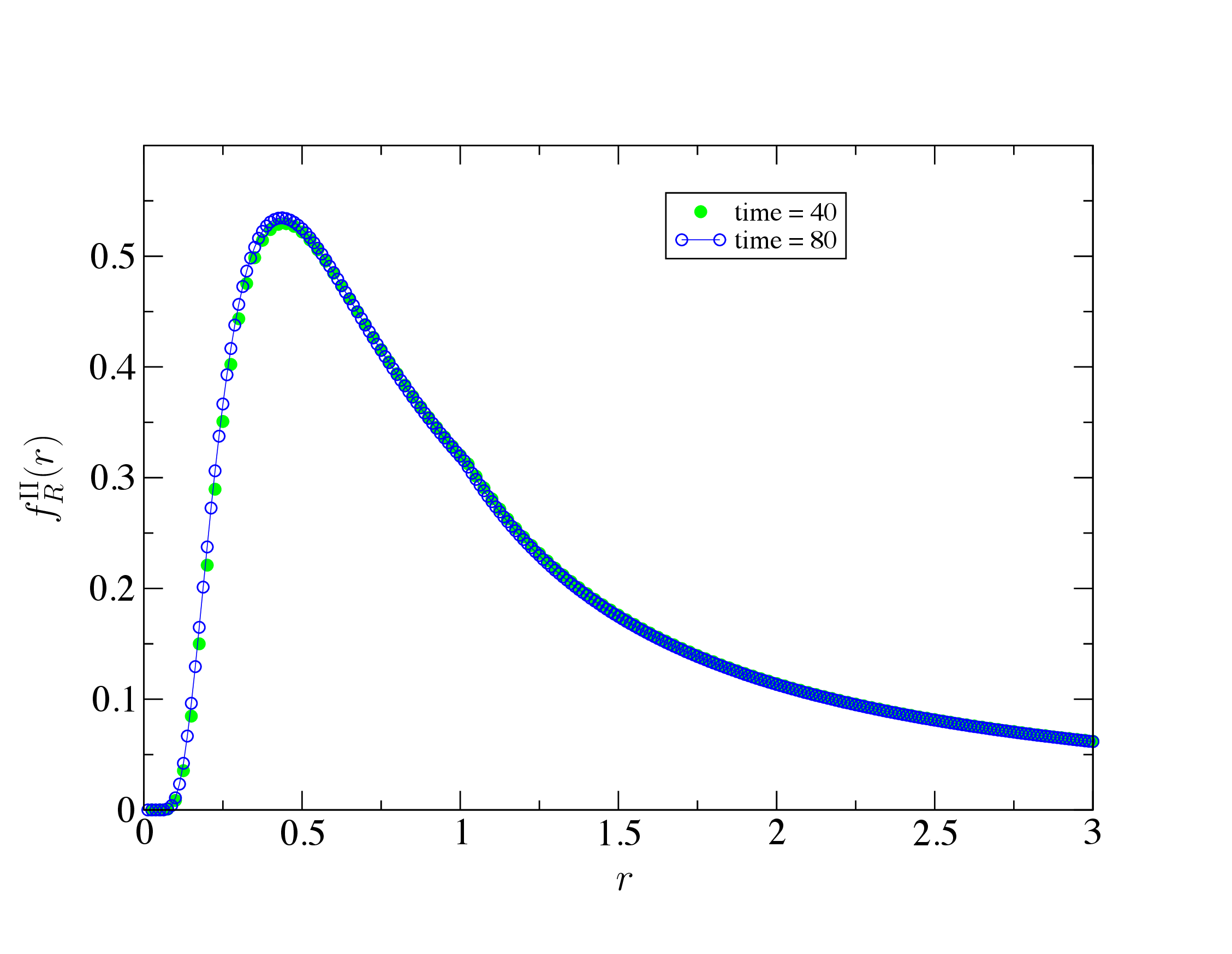}
\caption{\label{fig:fR_II}
Density $\fII_R(r)$ of the scaling variable $R$ for $\theta=1/2$, obtained by the same method as in figure~\ref{fig:fR_I}. 
The 
green full circles correspond to a random walk of 40 steps and the blue empty ones to a random walk of 80 steps. 
The line is a guide to the eyes connecting 
the blue circles. 
The tail has exponent $-3/2$, hence the first moment $\langle R\rangle$ of this distribution is infinite.
The discontinuity of the derivative at $r=1$ is clearly visible.}
\end{center}
\end{figure}

%
\section{Case II}
\label{II}
We now proceed to the sequence of intervals ${\cal C}^{\rm II}$, defined in~(\ref{def_conf}), following the line of reasoning employed for case I in the previous section.

\subsection{Distribution of $\LII(t)$}
We now consider
\beq
\LII(t)=\max(\tau_1,\ldots,\tau_{N+1}),
\eeq
whose distribution function,
\beq
\FII(t;\l)=\pro(\LII(t)\le\l),
\eeq
can be computed by the same method as for case I.
We first consider the joint distribution
\beqa
\FII_n(t;\l)=\pro(\LII(t)\le\l,N_t=n)
\nonumber\\
\fl=\int_{0}^{\l}{\rm d}\l_1\,\r(\l_1)\ldots\int_0^\l{\rm d}\l_n\,\r(\l_n)\int_0^\l{\rm d}\l_{n+1}\,\rho(\l_{n+1})
\,I\left(\sum_{i=1}^n\l_i<t<\sum_{i=1}^n\l_i+\l_{n+1}\right),
\eeqa
where $\{\l_1,\ldots,\l_n,\l_{n+1}\}$ is a realization of the configuration ${\cal C}^{\rm II}$, and
where $I(\cdot)=1$ if the condition inside the parentheses is satisfied and $I(\cdot)=0$ otherwise (see~(\ref{eq:fII})).
In Laplace space we have 
\beqa
\lap t \FII_n(t;\l)=\hat \FII_n(s;\l)
\nonumber\\
=
\left(\int_0^{\l}{\rm d}\tau\r(\tau)\e^{-s\tau}\right)^n
\int_0^\l{\rm d}\l_{n+1}\,\rho(\l_{n+1})\frac{1-\e^{-s \l_{n+1}}}{s} .
\eeqa
Thus
\beqa\label{eq:hatFII}
\fl\hat \FII(s;\l)=\sum_{n\ge0}\hat \FII_n(s;\l)
=\frac{1}{1-J(s;\l)}\int_0^\l{\rm d}\l_{n+1}\,\rho(\l_{n+1})\frac{1-\e^{-s \l_{n+1}}}{s},
\eeqa
and
\beqa\label{eq:FII}
\frac{1}{s}-\hat \FII(s;\l)
=\frac{1}{s}\frac{p_0(\l)}{1-J(s;\l)}
=\frac{1}{s}\,\frac{\e^{s\l}}{1+sI(s;\l)\e^{s\l}/p_0(\l)},
\eeqa
so that
\beq\label{eq:lmax_II}
\lap{t}\langle\LII(t)\rangle=\frac{1}{s}\,\int_0^\infty{\rm d}\l\,\frac{\e^{s\l}}{1+sI(s;\l)\e^{s\l}/p_0(\l)}.
\eeq
Normalization of the distribution of $\LII(t)$ can be checked on~(\ref{eq:hatFII}) by letting $\l\to\infty$.

Note that the expression~(\ref{eq:FII}) is simply related to~(\ref{eq:FI}).
By inversion, we obtain the relation
\beq\label{eq:relation}
\FII(t;\l)=\FI(t+\l,\l).
\eeq
\subsection{Probability of record breaking}
We now have
\beqa
\fl\QII(t)
&=&\pro(\LII(t)=\tau_{N+1}) =\pro(\tau_{N+1}>\max(\tau_1,\ldots,\tau_N))
\nonumber\\
\fl&=&\sum_{n\ge0}Q_n(t)=\sum_{n\ge0}\pro\big(\tau_{N+1}>\max(\tau_1,\ldots,\tau_n),N_t=n\big).
\eeqa
Explicitly, (setting $\l_{n+1}=y$, for short),
\beq
\fl\QII_n(t)=\int_0^\infty{\rm d}y\,\rho(y)\int_0^y{\rm d}\l_1\,\rho(\l_1)\ldots\int_0^y{\rm d}\l_n\,\rho(\l_n)\,
I\left(\sum_{i=1}^n\l_i<t<\sum_{i=1}^n\l_i+y\right).
\eeq
In Laplace space, after summing on $n$, we obtain
\beqa\label{eq:QII}
\fl \hat \QII(s)
=\frac{1}{s}\int_0^\infty{\rm d}y\,\frac{\rho(y)(1-\e^{-sy})}{1-\int_0^y{\rm d}\l\,\rho(\l)\e^{-sl}}
=\frac{1}{s}\int_0^\infty{\rm d}y\,\frac{\rho(y)(1-\e^{-sy})}{p_0(y)\e^{-sy}+sI(s;y)},
\eeqa
using~(\ref{eq:parties}).
As can be seen on~(\ref{eq:lmax_II}) and~(\ref{eq:QII}), there is no simple relationship, as in~(\ref{identity}), between $\LII(t)$ and $\QII(t)$ in the present case.

\subsection{Discussion}

We now review the behaviours of $\LII(t)$ and $\QII(t)$ according to the type of the distribution of intervals $\rho(\tau)$.

\smallskip\noindent{\it (i) Exponential distribution}

We have, from~(\ref{eq:FII}),
\beq\label{eq:hatFII+}
\frac{1}{s}-\hat\FII(s;l)=\frac{1+s}{s}\frac{\e^{s\l}}{1+s\,\e^{(1+s)\l}},
\eeq
from which no simple explicit expression of $\langle\LII(t)\rangle$ can be obtained by inversion.
Numerically one finds that $\langle\LII(t)\rangle$ is very close to $\langle\LI(t)\rangle$, i.e.,
\beq
\langle\LII(t)\rangle\approx\ln t+\gamma.
\eeq
Moreover, it is easy to show from~(\ref{eq:hatFII+}) that, for $s$ small and $\l$ large, one has
\begin{eqnarray}
\hat \FII(s;\l) \approx \hat F^{\rm I}(s;\l) \approx \frac{1}{s + \e^{-\l}} ,
\end{eqnarray}
which, as in~(\ref{eq:Gumbel}), shows that $(\ell^{\rm II}_{\max}(t) - \ln t)$ is distributed according to a Gumbel distribution.

On the other hand we find that 
\beq
\hat \QII (s)=\frac{1+s}{s}\int_0^\infty{\rm d}y\,\frac{\e^{sy}-1}{1+s\, \e^{(1+s)y}},
\eeq
from which, again, no simple explicit expression of $\QII(t)$ can be obtained. 
For $s$ small, we have
$\hat \QII (s) \approx (1/2) (\ln s)^2$, yielding
\beq
\QII(t)\approx\frac{\ln t}{t} ,
\eeq
which is fully compatible with numerical simulations.

\smallskip\noindent{\it (ii) Broad distribution with index $0<\theta<1$}

It turns out, as explained below, that the first moment of $\LII(t)$ is not defined.
In contrast, $\QII(t)$ has a well defined limit at large times.
This probability was investigated previously by Scheffer~\cite{scheffer} as the probability that the last excursion of Brownian motion containing $t$ is the longest\footnote{
``Let $x_t$ be the duration of the excursion from 0 straddling $t$. 
Then it is asked to compute the probability that this present excursion has a record duration, i.e., that $x_t$ is greater than the maximum of the durations of all previous excursions."~\cite{scheffer}}.

The aim of this section is to recover the result of Scheffer by simple methods, as already done in~\cite{us1,us2}, and to extend it to any value of $0<\theta<1$.
We will also determine the density of $\LII(t)$ in the present case.
Using the same methods as for case I, we have, in the scaling regime, starting from~(\ref{eq:FII}):
\beq\label{eq:basisII}
\fl\frac{1}{s}-\hat \FII(s;\l) \approx \frac{1}{s} \hat f^{\rm II}_V(s \ell) , \qquad \hat f^{\rm II}_V(x) = \frac{\e^{x}}{1+x^\theta\e^x\int_0^x{\rm d}u\,u^{-\theta}\e^{-u}}.
\eeq
As for case I, the random variables $R_t$ and $V_t$ (both now defined between 0 and $\infty$) have limiting distributions in this regime, denoted by $\fII_R(r)$ and $\fII_V(v)$, and the scaling function $\hat f^{\rm II}_V(x)$~(\ref{eq:basisII}) is the Laplace transform with respect to $v$ of the latter.
In the particular case where $\theta=1/2$,
\beq
\hat f^{\rm II}_V(x)
=\frac{1}{\e^{-x}+\sqrt{\pi x}\erf\sqrt{x}}
\approx \frac{1}{\sqrt{\pi x}}.
\eeq
The integral of this expression diverges, hence $\langle\LII\rangle/t$ diverges.
The explanation of this result is that the largest interval of the sequence ${\cal C^{\rm II}}$ has a finite probability of being the last one, $\tau_{N+1}$, which itself does not possess a finite first moment.
This also holds for any $0<\theta<1$. 

Comparing~(\ref{eq:basisI}) and (\ref{eq:basisII}) we notice that $\hat \fII_V(x)=\e^{x}\hat \fI_V(x)$, hence we have the relation
$\fII_V(v)=\fI_V(v+1)$ (which can also be deduced directly from~(\ref{eq:relation})). 
This implies the relation $f^{\rm II}_R(r) = (r+1)^{-2} f_R^{\rm I}(r/(r+1))$. 
Hence the asymptotic  behaviours of $f_V^{\rm II}(v)$ and $f_R^{\rm II}(r)$ follow straightforwardly from~(\ref{eq:fv_smallv}), (\ref{eq:fv_largev}) and (\ref{eq:asympt_f_RI}). 
Besides, as for case I, the first moments of $V$ have simple explicit expressions (see~(\ref{eq:moments_V_I})).
More generally we have, with $p$ an integer,
\beq
\langle (V^{\rm II})^p\rangle=\langle (V^{\rm I}-1)^p\rangle.
\eeq

On the other hand,
\beq\label{gen_scheffer}
\fl\QII_\infty=\lim_{t \to \infty} Q^{\rm II}(t)=\lim_{s\to0}s\hat \QII(s)=\theta\int_0^\infty{\rm d}x\,
\frac{(\e^{x}-1)/x}{1+x^\theta\e^x\int_0^x{\rm d}u\,u^{-\theta}\e^{-u}}
\eeq
is finite.
For $\theta=1/2$, one recovers the result of Scheffer~\cite{scheffer}, 
\beq\label{eq:scheffer}
\QII_\infty= \frac{1}{2} \int_0^\infty {\rm d}x\, \frac{\e^x-1}{x + \sqrt{\pi }x^{3/2} \, \e^x \, \erf\sqrt{x}} = 0.800\,310 \dots\,.
\eeq

\smallskip\noindent{\it (iii) Broad distribution with index $1<\theta<2$}

Performing the same analysis as in section~\ref{sec:lev_I_tet>1}, we conclude that, for $1<\theta<2$, comparing~(\ref{eq:FI}) and (\ref{eq:FII}), we have
\beq
\frac{1}{s}-\hat F^{\rm II}(s;\l)\approx\frac{1}{s}-\hat F^{\rm I}(s;\l)
\approx \frac{1}{s}\frac{1}{1+s\langle\tau\rangle(\l/\tau_0)^\theta},
\eeq
hence the two random variables $\LI(t)$ and $\LII(t)$
have asymptotically the same statistics.
Restarting from~(\ref{eq:QII}), we have 
\beq
\hat \QII(s)=\frac{1}{s}\int_0^y
{\rm d}y\,\frac{\rho(y)(1-\e^{-sy})}{\tau_0^\theta y^{-\theta}\e^{-sy}+s\langle\tau\rangle},
\eeq
i.e., using the analysis performed in section~\ref{sec:lev_I_tet>1} (see~(\ref{eq:FI>1})), 
\beq
\hat \QII(s)=\theta\int_0^\infty{\rm d}y\,\frac{1}{1+s\langle\tau\rangle(y/\tau_0)^\theta}
=\theta\,\hat\QI(s),
\eeq
hence $\QII(t)=\theta\,\QI(t)$, where $\QI(t)$ is given in~(\ref{Q_I_levy2}).

\begin{figure}
\begin{center}
\includegraphics[angle=0,width=0.9\linewidth]{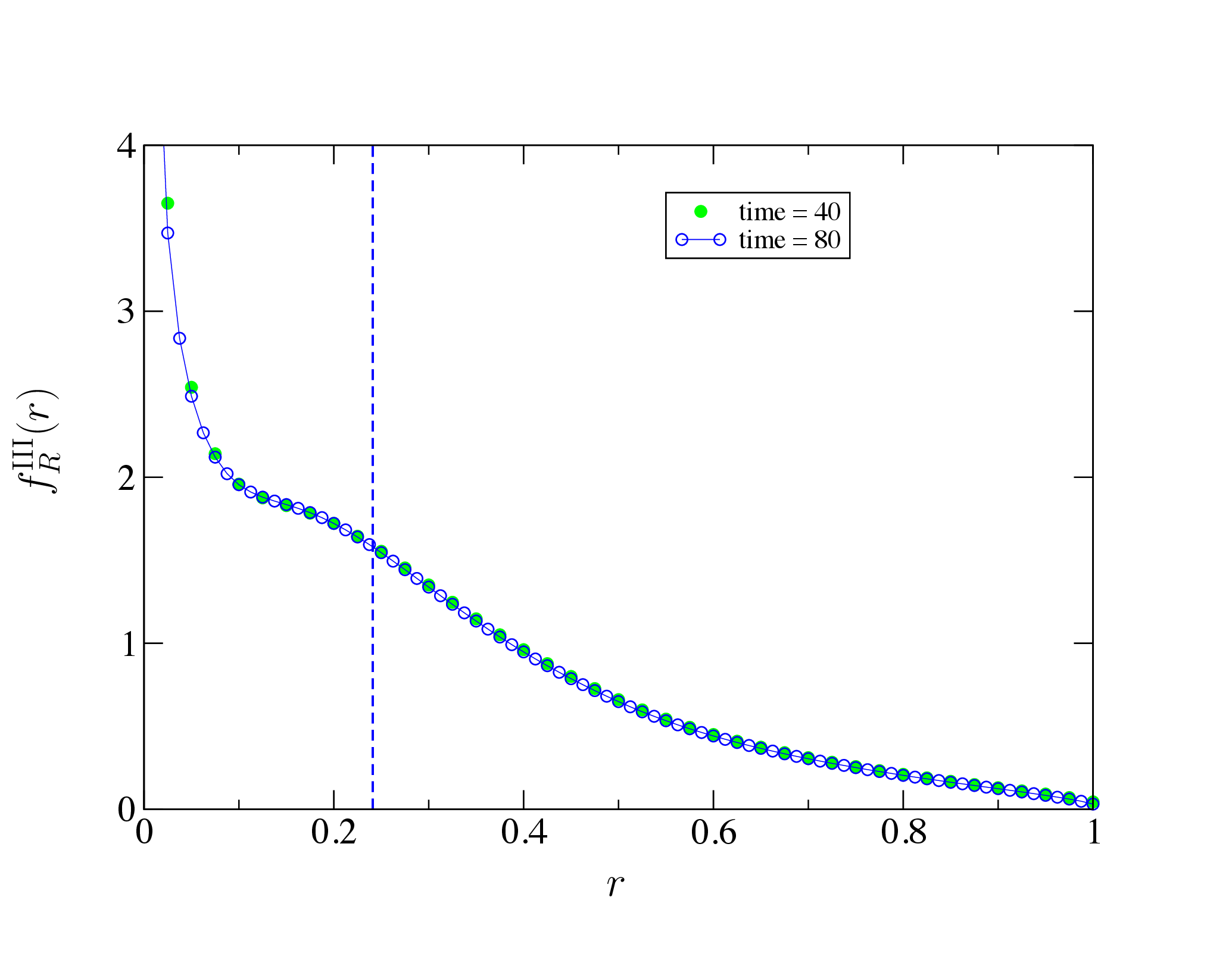}
\caption{\label{fig:fR_III}
Density $\fIII_R(r)$ of the scaling variable $R$ for $\theta=1/2$, obtained by the same method as in figures~\ref{fig:fR_I} and~\ref{fig:fR_II}.
\textcolor{black}{The green full circles correspond to a random walk of 40 steps and the blue empty ones to a random walk of 80 steps. 
The line is a guide to the eyes connecting 
the blue circles.}
The vertical dotted line indicates the value of the first moment $\langle R\rangle\approx 0.241\,749\dots$ given in~(\ref{eq:result_III}).
}
\end{center}
\end{figure}

\begin{figure}
\begin{center}
\includegraphics[angle=0,width=0.9\linewidth]{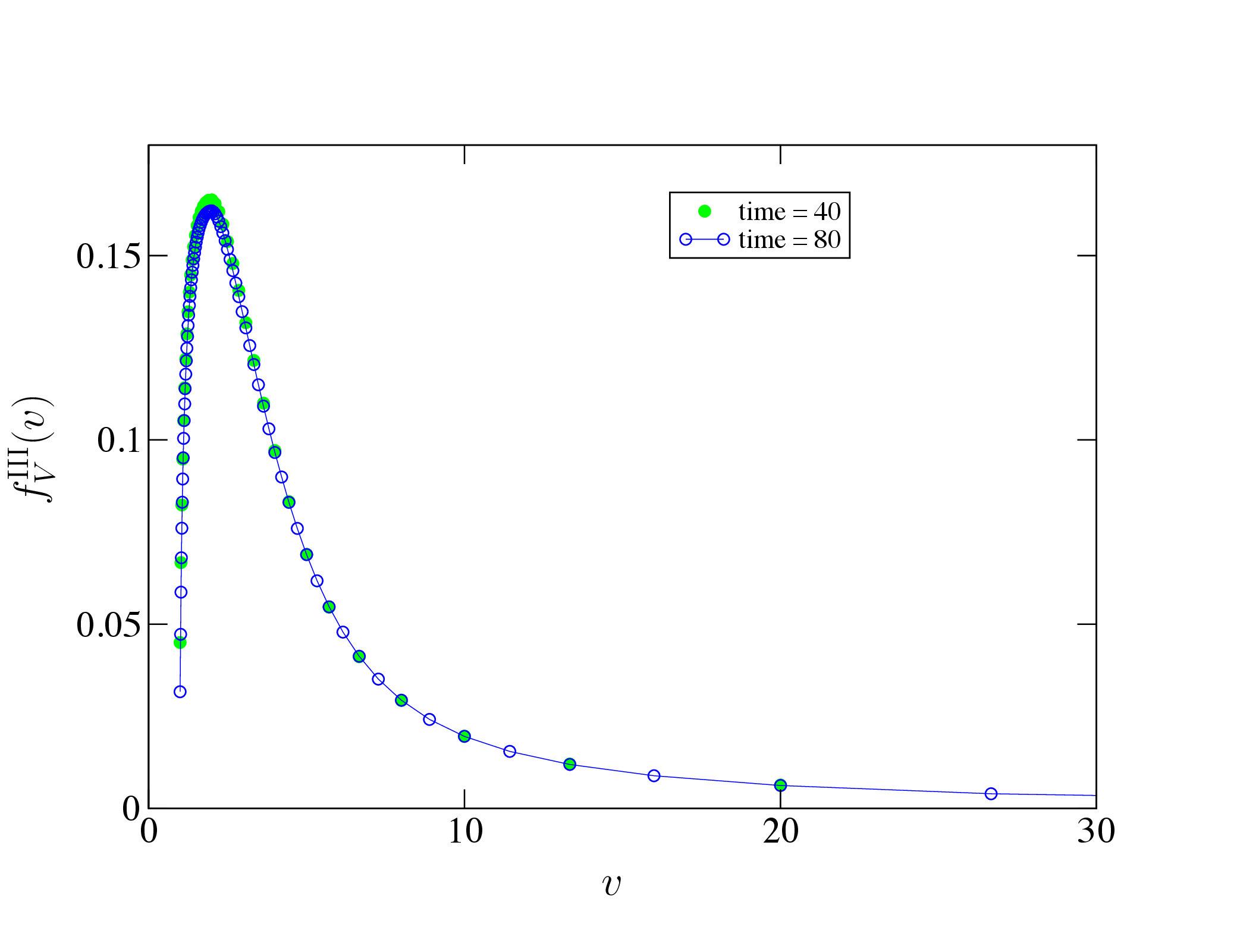}
\caption{\label{fig:fV_III}
Probability density $\fIII_V(v)$, where $V=t/\LIII(t)$ as $t\to\infty$ for $\theta=1/2$, obtained from the data of figure~\ref{fig:fR_III}, using $V_t=1/R_t$.
The function is defined between $v=1$ and infinity and vanishes as $\sqrt{v-1}$ close to $v=1$.}
\end{center}
\end{figure}

%
\section{Case III}
\label{III}

We finally consider the sequence of intervals ${\cal C}^{\rm III}$, defined in~(\ref{def_conf}), following the line of reasoning employed for the two previous cases, I and II, for the determination of the statistics of the longest interval occurring between $0$ and $t$.

\subsection{Distribution of $\LIII(t)$}
By definition,
\beq
\LIII(t)=\max(\tau_1,\ldots,\tau_{N}).
\eeq
We adopt the convention that if there is no event between 0 and $t$, i.e., $N_t=0$, then $\LIII=0$.
Its distribution function
\beq
\FIII(t;\l)=\pro(\LIII(t)\le\l)
\eeq
can be computed by the same method as for cases I and II.
We have (see~(\ref{eq:fIII}))
\beqa
\FIII_n(t;\l)=\int_0^\l {\rm d}\l_1\ldots\int_0^\l{\rm d}\l_n\, \fIII(t;\l_1,\ldots,\l_n,n)\\
\fl=\int_0^\l {\rm d}\l_1\ldots\int_0^\l{\rm d}\l_n\,
\r(\l_1)\ldots\r(\l_n)\,\int_0^\infty{\rm d}a\,p_0(a)\,\delta\left(\sum_{i=1}^n\l_i+a-t\right),
\eeqa
where $\{\l_1,\ldots,\l_n\}$ is a realization of the configuration ${\cal C}^{\rm III}$,
yielding, in Laplace space, after summation on $n$, 
\beq\label{eq:hatFIII}
\hat \FIII(s;\l)=\frac{1-\hat\rho(s)}{s}\frac{1}{1-J(s;\l)},
\eeq
hence
\beq\label{eq:lmaxIII}
\frac{1}{s}-\hat \FIII(s;\l)=\frac{1}{s}\frac{\hat\rho(s)-J(s;\l)}{1-J(s;\l)}.
\eeq
Normalization of the distribution of $\LIII(t)$ can be checked on~(\ref{eq:hatFIII}). 

\subsection{Probability of record breaking $\QIII(t)$}

There are two possible ways of computing this probability. 
The first one proceeds as for the two previous cases, I and II.
We have
\beqa
\fl\QIII(t)
&=&\pro(\LIII(t)=\tau_{N}),
=\pro(\tau_{N}>\max(\tau_1,\ldots,\tau_{N-1})),
\nonumber\\
\fl&=&\sum_{n\ge0}Q_n(t)=\sum_{n\ge0}\pro\big(\tau_{N}>\max(\tau_1,\ldots,\tau_{N-1}),N_t=n\big).
\eeqa
Explicitly,
\beq
\fl\QIII_n(t)=\int_0^\infty{\rm d}\l_{n}\,\int_0^{\l_n}{\rm d}\l_1\,\ldots\int_0^{\l_n}{\rm d}\l_{n-1}\,
f^{{\rm III}}(t;\l_1,\dots,\l_{n-1},\l_{n},n).
\eeq
In Laplace space, after summing on $n$, we obtain
\beqa
\hat \QIII(s)
&=&\hat p_0(s)\int_0^\infty{\rm d}y\,\frac{\rho(y)\e^{-sy}}{1-\int_0^y{\rm d}\l\,\rho(\l)\e^{-sl}},
\nonumber\\
&=&\hat p_0(s)\int_0^\infty{\rm d}y\,\frac{{\rm d}J(s;y)/{\rm d}y}{1-J(s;y)},
\eeqa
where $J(s;y)=\int_0^y{\rm d}\l\,\rho(\l)\e^{-s\l}$.
Finally,
\beq\label{eq:QIII0}
\hat \QIII(s)=\hat p_0(s)\int_0^{\hat\rho(s)}\frac{{\rm d}J}{1-J}
=-\frac{1-\hat\rho(s)}{s}\ln(1-\hat\rho(s)).
\eeq

We now show that the same result can be recovered by a more intuitive method, which relies on the idea that the $N_t$ intervals $\tau_1,\ldots,\tau_{N}$ are expected to play the same role.
One is therefore tempted to apply the well known fact, valid for i.i.d. random variables, that the probability for the last variable to be the largest, that is to say, the probability of record breaking, is equal to the inverse number of random variables.
We are thus led to write
\beq
\QIII_n(t)=\pro(\tau_n>\max(\tau_1,\ldots,\tau_{n-1}))=\frac{p_n}{n},\quad (n>0),
\eeq
where $p_n=\pro(N_t=n)$, thus
\beq
\fl\QIII(t)=\pro(\tau_N>\max(\tau_1,\ldots,\tau_{N-1}), N_t=n)=
\sum_{n\ge1}\QIII_n(t)=\left\langle\frac{1}{N_t}\right\rangle,
\eeq
(without $N_t=0$).
In Laplace space,
\beqa
\hat \QIII(s)&=&\sum_{n\ge1}\frac{\hat p_n(s)}{n}
=\frac{1-\hat\rho(s)}{s}\sum_{n\ge1}\frac{\hat \rho(s)^n}{n},
\nonumber\\ 
&=&-\frac{1-\hat\rho(s)}{s}\ln(1-\hat\rho(s)),
\eeqa
which is~(\ref{eq:QIII0}) above.

\subsection{Discussion}

\smallskip\noindent{\it (i) Exponential distribution of intervals}

Using~(\ref{eq:lmaxIII}) and~(\ref{eq:QIII0}), we find
\beq
\lap{t}\langle\LIII(t)\rangle=\frac{1}{s}\int_0^\infty{\rm d}\l\frac{1}{1+s\,\e^{(s+1)\l}}
=\frac{1}{s(1+s)}\ln\left(1+\frac{1}{s}\right),
\eeq
and
\beq
\hat \QIII(s)=
\frac{1}{1+s}\ln\left(1+\frac{1}{s}\right).
\eeq
From these two expressions and from~(\ref{eq:expI}) it follows that 
\beqa
\langle\LIII(t)\rangle
&=&\langle\LI(t)\rangle-\QIII(t),
\nonumber\\
&\approx&\ln t+\gamma,
\eeqa
where
\beq\label{eq:QIII_exp}
\QIII(t)=-\e^{-t}\int_0^t{\rm d}u\,\frac{1-\e^{u}}{u}
\approx\frac{1}{t}.
\eeq
In practice $\langle\LI(t)\rangle$ and $\langle\LIII(t)\rangle$, as well as $\QI(t)$ and $\QIII(t)$, are numerically rapidly indistinguishable in this exponential case.

One can also check from~(\ref{eq:lmaxIII}) that
\beq
\hat \FIII(s;\l) \approx \hat \FI(s;\l) \approx \frac{1}{s + \e^{-\l}} ,
\eeq
implying, as in~(\ref{eq:Gumbel}), that $(\ell^{\rm III}_{\max}(t) - \ln t)$ is distributed according to a Gumbel distribution.

\smallskip\noindent{\it (ii) Broad distribution of intervals with index $0<\theta<1$}

The starting point of our analysis is~(\ref{eq:lmaxIII}). 
By inserting the small $s$  behaviour (\ref{ro_broad}) of $\hat \rho(s)$ and ~(\ref{eq:scaling_IJ_1}) of $J(s;l)$ into~(\ref{eq:lmaxIII}), one obtains, in the limit $s \to 0$, $\ell \to \infty$, keeping $x = s \ell$ fixed:
\beq\label{eq:basisIII}
\fl\frac{1}{s}-\hat \FIII(s;\l)\approx\frac{1}{s} \hat f^{\rm III}_V(s \ell) , \qquad \hat f^{\rm III}_V(x) = 
\frac{1-x^\theta\e^x\int_x^\infty{\rm d}u\,u^{-\theta}\e^{-u}}
{1+x^\theta\e^x\int_0^x{\rm d}u\,u^{-\theta}\e^{-u}} .
\eeq
As for cases I and II, the scaling function $\hat f^{\rm III}_V(x)$ is the Laplace transform with respect to $v$ of the limiting distribution $\fIII_V(v)$ of the rescaled variable $V_t = t/\ell^{\rm III}_{\max}(t)$ (see~(\ref{eq:fvlap})). 
Besides, from~(\ref{eq:basisIII}) we get
\beq
\fl\lim_{t\to\infty}\frac{1}{t}\langle\LIII(t)\rangle=\int_0^\infty{\rm d}x\,\hat f^{\rm III}_V(x)
=\int_0^\infty{\rm d}x\,
\frac{1-x^\theta\e^x\int_x^\infty{\rm d}u\,u^{-\theta}\e^{-u}}
{1+x^\theta\e^x\int_0^x{\rm d}u\,u^{-\theta}\e^{-u}}.
\eeq
In particular, for $\theta=1/2$ this yields 
\beq\label{eq:result_III}
\lim_{t\to\infty}\frac{1}{t}\langle\LIII(t)\rangle=\int_0^\infty{\rm d}x\,
\frac{1-\sqrt{\pi x\,}\e^{x}\,{\rm erfc\,}\sqrt{x}}{1+\sqrt{\pi x}\,\e^{x}\erf\sqrt{x}}=0.241\,749\dots.
\eeq

Let us briefly discuss the  behaviours of the densities $f_R^{\rm III}(r)$ and $f_V^{\rm III}(v)$. 
As for case I (see, e.g.,~(\ref{eq:fv_smallv})), one expects that $f_V^{\rm III}(v)$ has different expressions on each interval $[k,k+1]$ where $k=1,2, \ldots$, with singularities at $v= 2,3,\ldots$. 
On the other hand, the asymptotic  behaviour of $f_V^{\rm III}(v)$ can be easily obtained from its Laplace transform $\hat f_V^{\rm III}(x)$. 
In particular, for large $x$, one has $\hat f_V^{\rm III}(x) \sim (\theta/\Gamma(1-\theta)) x^{-1 - \theta} \e^{-x}$ from which it follows that
\begin{eqnarray}\label{eq:fIII_small_v}
f_V^{\rm III}(v) \approx \frac{\sin{\pi \theta}}{\pi} (v-1)^{\theta} , \; v \to 1 .
\end{eqnarray}

The large $v$  behaviour of $f_V^{\rm III}(v)$ is dominated by the non-analytic  behaviour of $\hat f_V^{\rm III}(x)$ close to the origin, while the pole of $\hat f_V^{\rm III}(x)$ yields here subleading corrections. 
One has indeed $\hat f_V^{\rm III}(x) \approx 1 - x^{\theta} \Gamma(1-\theta)$, when $x \to 0$, implying
\begin{eqnarray}\label{eq:fIII_large_v}
f_V^{\rm III}(v) \approx \theta\, v^{-1-\theta} , \; v \to \infty ,
\end{eqnarray}
which shows that $V$ has no finite moments. 
From the asymptotic  behaviours of $f_{V}^{\rm III}(v)$ in~(\ref{eq:fIII_small_v}) and~(\ref{eq:fIII_large_v}) one obtains those of $f_R^{\rm III}(r)$ as
\begin{eqnarray}\label{eq:asympt_f_RIII}
f^{\rm III}_R(r) \approx
\left\{
 \begin{array}{lll}
 \theta\, r^{\theta - 1} , & \; r \to 0 \\
 & \\
 (\sin(\pi\theta)/\pi) (1-r)^\theta , & \; r \to 1 .
 \end{array}
\right.
\end{eqnarray} 
Figure~\ref{fig:fR_III} depicts the density $\fIII_R(r)$ of the scaling variable $R$, for $\theta=1/2$, using the same method as in figures~\ref{fig:fR_I} and~\ref{fig:fR_II}.
Figure~\ref{fig:fV_III} is deduced from figure~\ref{fig:fR_III}.

\begin{figure}
\begin{center}
\includegraphics[angle=0,width=0.9\linewidth]{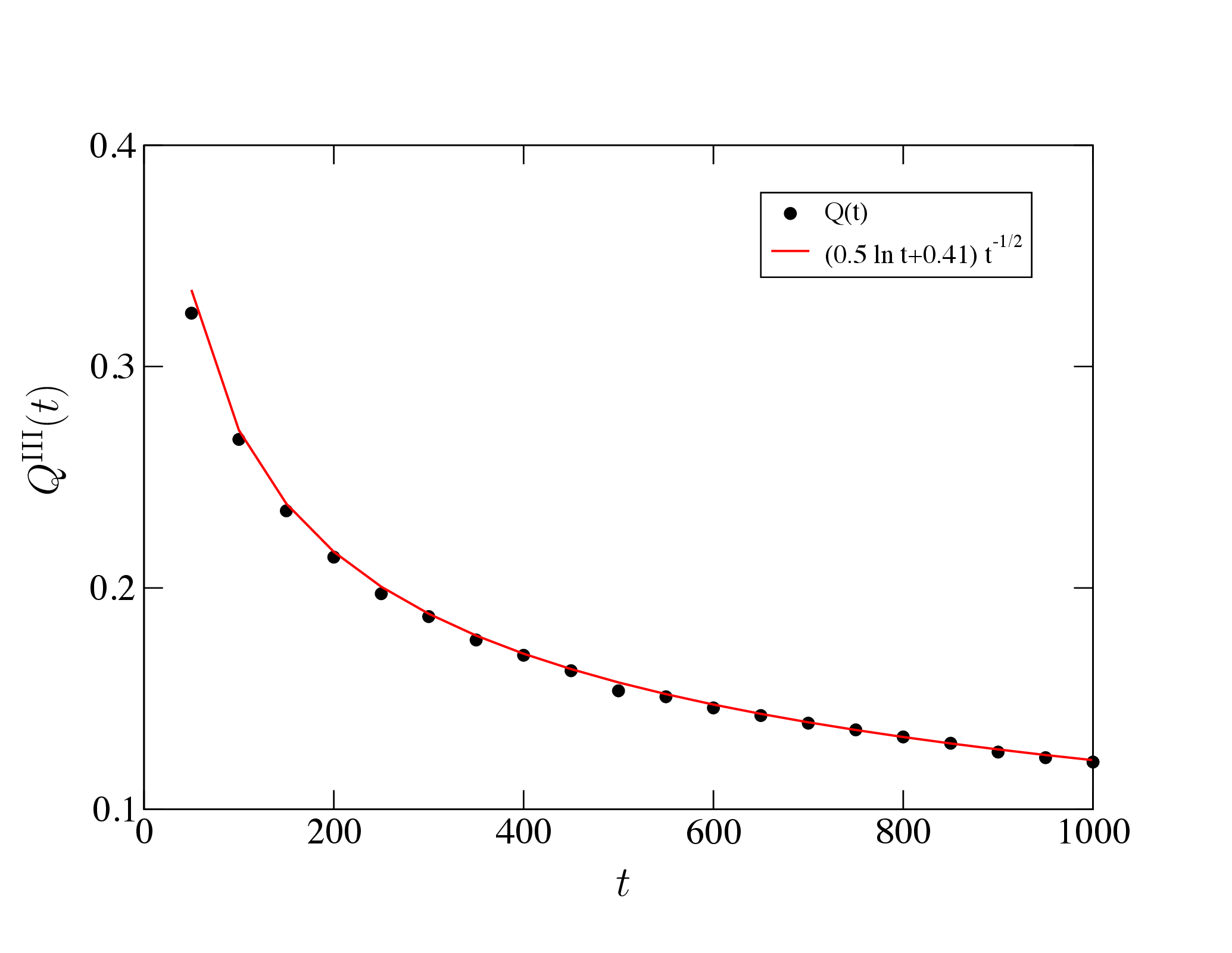}
\caption{\label{fig:QIII}
\textcolor{black}{Comparison of the analytical prediction~(\ref{eq:QIII_lev}) (red solid line) for 
$Q^{\rm III}(t)$ with simulations (black full circles) with $\theta=1/2$.}}
\end{center}
\end{figure}

Next, we have
\beq
\hat \QIII(s)\approx
-a s^{\theta-1}\ln (a s^{\theta}),
\eeq
hence
\beq\label{eq:QIII_lev}
\QIII(t)\approx
(A\ln t+B)\,t^{-\theta},
\eeq
with $A=\theta\tau_0^\theta$, and $B$ is a constant depending on $\tau_0$ and $\theta$.
For instance, taking a distribution of intervals $\rho(\tau)$ generated by $U^{-1/\theta}$ where $U$ is uniform between 0 and 1, then $\tau_0=1$, and if for instance $\theta=1/2$, we have $B=\gamma/2+\ln (2/\pi)$ (see figure~\ref{fig:QIII}).

%
\smallskip\noindent{\it (iii) Broad distribution of intervals with index $1<\theta<2$}

Again we find
\beq
\frac{1}{s}-\hat F^{\rm III}(s;\l)\approx\frac{1}{s}-\hat F^{\rm I}(s;\l),
\eeq
hence the distribution of $\LIII(t)$ is asymptotically identical to that of $\LI(t)$.

Finally, a simple analysis yields
\beq
\QIII(t)=\left\langle\frac{1}{N_t}\right\rangle\approx\frac{\langle\tau\rangle}{t}.
\eeq
This behaviour is akin to the case of i.i.d. random variables where $Q(n)=1/n$,
since here we have
$\langle N_t\rangle\approx t/\langle\tau\rangle$.

\section{Conclusion and discussion}
\label{sec:discuss}

In this paper, we have investigated the statistics of the longest interval in renewal processes. 
Let us put our results in the context of recent works starting with~\cite{ziff}, where the authors studied, among others, the longest lasting record in symmetric random walks. 
Thanks to the discrete renewal properties of the records of random walks \cite{feller,ziff}, the longest lasting record corresponds precisely to the longest interval in a renewal process with a distribution of intervals $\rho(\tau) \sim \tau^{-1-\theta}$, where $\theta = 1/2$. 
Ref.~\cite{ziff} considered the situation referred to as case I in the present work (see~(\ref{def_conf})) and obtained the rescaled first moment $\langle \ell^{\rm I}_{\max}(t)\rangle/t$. 
A subsequent paper focused on the related question of $\langle \ell^{\rm I}_{\max}(t)\rangle/t$ for a broader class of continuous renewal processes, where $\rho(\tau) \sim \tau^{-1-\theta}$, with $0 < \theta < 2$, and discussed the relevance of this quantity for stochastic processes in nonequilibrium systems~\cite{us1}. 
This reference also studied the probability of record breaking $Q^{\rm I}(t)$ and discussed briefly the possible extensions of these quantities, $\ell^{\rm I}_{\max}(t)$ and $Q^{\rm I}(t)$, to different ensembles denoted by ${\cal C}^{\rm II}$ and ${\cal C}^{\rm III}$ in the present paper (see~(\ref{def_conf})). 
Then, coming back to the record statistics of random walks, a detailed analysis of the statistics of  $\ell^{\rm \alpha}_{\max} (t)$ and $Q^{\rm \alpha}(t)$ for the three cases $\alpha=$ I, II, III was recently performed~\cite{us2}. 

The present study of the statistics of $\ell_{\max}^{\alpha}(t)$ 
and of $Q^{\alpha}(t)$ for a large panel of distributions of intervals $\rho(\tau)$ completes the previous aforementioned works. 
We found a rich variety of  behaviours, depending on the tail of $\rho(\tau)$, which we briefly summarize here (see also tables \ref{tab:exp}, \ref{tab:levy<2} and \ref{tab:levy<1}).
Concerning the statistics of $\ell_{\max}^{\alpha}(t)$, we showed that, if $\rho(\tau)$ decays faster than $1/\tau^2$, the fluctuations of $\ell_{\max}^{\alpha}(t)$ are given in the large $t$ limit by the classical theory of extreme value statistics for i.i.d. random variables. 
Indeed the distribution of $\ell_{\max}^{\alpha}(t)$, properly shifted and scaled, converges in the large $t$ limit to one of the standard distributions of extreme value statistics for i.i.d. random variables, namely Gumbel, Fr\'echet or Weibull, depending on the tail of $\rho(\tau)$, and provided that $\tau^2 \rho(\tau) \to 0$ when $\tau \to \infty$. 
This shows in particular that the global constraint imposing that the sum of the time intervals is fixed to $t$ becomes irrelevant, in this case, for large $t$. 
This is however not the case if $\rho(\tau)$ has a power-law decay with tail exponent $0<\theta<1$, where the limiting distribution of the rescaled variable $R_t=\ell^{\alpha}_{\max}(t)/t$, computed in the various cases $\alpha =$ I, II and III, is non-trivial. 
A generic feature of these limiting distributions $f_R^{\alpha}(r)$ is that they exhibit non-analyticities at values $r = 1/k$ with $k = 2,3,\ldots$ for the cases $\alpha =$ I, III and $k = 1,2,\ldots$ for $\alpha =$ II. 
Such non-analytic densities have been obtained previously in several related, though different, situations~\cite{Lam61,Wen64,DF87,FIK95,KE94,EK97,glrecord}. 
We refer the reader to figures~\ref{fig:fR_I}, \ref{fig:fR_II} and \ref{fig:fR_III} for a plot of these densities in the case $\theta = 1/2$, corresponding to the renewal sequence of zeros or records of random walks. 
Although the variable $R_t=\ell^{\alpha}_{\max}(t)/t$ seems physically more natural, it turns out that the distribution of $V_t = 1/R_t = t/\ell^\alpha_{\max}(t)$ is easier to study. 
In case I, the limiting distribution of $V_t$ had been computed before by Lamperti in \cite{Lam61}. 
In this case, our results provide a simple alternative derivation of his result. 
As mentioned in section~\ref{sec:model}, $f^{\rm I}_V(v)$ is identical to the distribution found by Darling for the ratio between the maximum term and the sum of heavy tailed random variables with $0<\theta<1$~\cite{pitman}. 
We refer the reader to figures~\ref{fig:fV_I} and \ref{fig:fV_III} for plots of the limiting distributions $f_V^{\rm I}(v)$ and $f_V^{\rm III}(v)$ for $\theta=1/2$ (while $f_V^{\rm II}(v) = f_V^{\rm I}(v+1)$). 
A nice feature of these distributions in the case $0<\theta < 1$ is that they are completely universal, depending only on the exponent $\theta$ and not on the ``microscopic'' details of the distribution $\rho(\tau)$.

We also found a rich  behaviour of the probability of record breaking $Q^{\rm \alpha}(t)$ depending on the time distribution $\rho(\tau)$. In particular, we showed that in most cases $Q^{\rm \alpha}(t)$ behaves quite differently for renewal processes and for i.i.d. random variables. 
Again, the case $\theta < 1$ is particularly interesting as it gives rise to universal constants, for cases I and II. 
In particular, we recovered in a simple way a result obtained by Scheffer \cite{scheffer} for $\QII_\infty$ with $\theta =1/2$, which we have generalized to any value of $\theta$ (see~(\ref{gen_scheffer})).

It is remarkable that the observables $\ell_{\max}^\alpha(t)$ and $Q^\alpha(t)$ are quite sensitive to the last excursion, as the cases {\rm I, II} and ${\rm III}$ show different  behaviours (see tables \ref{tab:exp}, \ref{tab:levy<2} and \ref{tab:levy<1}). And it will be interesting to extend this study to other observables, like for instance the smallest interval in renewal processes, whose average values were computed in some cases in the context of record statistics \cite{ziff,MSW2012,us2}.

The influence of the last interval was also recently pointed out in the context
of anomalous diffusion~\cite{barkai1,barkai2}\footnote{We are indebted to Eli Barkai for pointing out refs.~\cite{barkai1,barkai2} to us.}.

\ack
 SNM and GS acknowledge support by ANR grant
2011-BS04-013-01 WALKMAT and in part by the Indo-French Centre for the Promotion of Advanced Research under Project 4604-3. GS acknowledges support from Labex-PALM (Project Randmat).

\section*{Appendix}

\subsection{Notations}

Our notations are those commonly used in probability theory: 
if $X$ is a random variable, then its distribution function reads
\beq
F_X(x)=\pro(X<x),
\eeq
and its density is 
\beq
f_X(x)=\frac{{\rm d}F_X(x)}{{\rm d}x}.
\eeq
Likewise, for several random variables, we have
\beq
F_{X_1,X_2,\ldots}(x_1,x_2,\ldots)=\pro(X_1<x_1,X_2<x_2,\ldots),
\eeq
with the associated density $f_{X_1,X_2,\ldots}(x_1,x_2,\ldots)$.
When permitted by the context, we will omit the variables in subscript.

\subsection{Joint probability densities}

\subsubsection*{{\rm (I)}}

The joint probability density of $\tau_1,\ldots,\tau_N,A_t,N_t$ is
\beqa
f_{\tau_1,\ldots,\tau_N,A_t,N_t}(t;\l_1,\ldots,\l_n,a,n)
\nonumber\\
=\langle\delta(\tau_1-\l_1)\ldots\delta(\tau_n-\l_n)
\delta(A_t-a)I(t_n<t<t_{n+1})\rangle,
\eeqa
where $I(\cdot)=1$ if the condition inside the parentheses is satisfied and $I(\cdot)=0$ otherwise.
This yields
\beqa
f_{\tau_1,\ldots,\tau_N,A_t,N_t}(t;\l_1,\ldots,\l_n,a,n)
\nonumber\\
=\r(\l_1)\ldots\r(\l_n)\,p_0(a)\,\delta\big(\sum_{i=1}^n\l_i+a-t\big).
\label{eq:fI}
\eeqa
Its Laplace transform with respect to time reads
\beqa\label{eq:applapI}
\fl\lap {t}f_{\tau_1,\ldots,\tau_N,A_t,N_t}(t;\l_1,\ldots,\l_n,a,n)
=\hat f_{\tau_1,\ldots,\tau_N,A_t,N_t}(s;\l_1,\ldots,\l_n,a,n)
\nonumber\\
=\r(\l_1)\ldots\r(\l_n)\,\e^{-s\sum_i\l_i}
p_0(a)\e^{-s a}.
\eeqa
Laplace transforming with respect to age yields
\beqa\label{eq:applapI+}
\fl\hat f_{\tau_1,\ldots,\tau_N,A_t,N_t}(s;\l_1,\ldots,\l_n,u,n)
=\r(\l_1)\ldots\r(\l_n)\,\e^{-s\sum_i\l_i}
\frac{1-\hat\r(s+u)}{s+u}.
\eeqa

\subsubsection*{{\rm (II)}}

Likewise, the joint probability density of $\tau_1,\ldots,\tau_{N+1},N_t$ is
\beqa
f_{\tau_1,\ldots,\tau_{N+1},N_t}(t;\l_1,\ldots,\l_{n+1},n)
\nonumber\\
=\langle\delta(\tau_1-\l_1)\ldots\delta(\tau_{n+1}-\l_{n+1})
I(t_n<t<t_{n+1})\rangle,
\eeqa
yielding
\beqa
\fl f_{\tau_1,\ldots,\tau_{N+1},N_t}(t;\l_1,\ldots,\l_{n+1},n)
=\r(\l_1)\ldots\r(\l_{n+1})\,I(t_n<t<t_{n}+\l_{n+1}).
\label{eq:fII}
\eeqa
In Laplace space with respect to $t$, we have
\beqa\label{eq:applapII}
\fl\lap {t}f_{\tau_1,\ldots,\tau_{N+1},N_t}(t;\l_1,\ldots,\l_{n+1},n)
=\hat f_{\tau_1,\ldots,\tau_{N+1},N_t}(s;\l_1,\ldots,\l_n,\l_{n+1},n)
\nonumber\\
=\r(\l_1)\ldots\r(\l_n)\,\e^{-s\sum_i\l_i}
\rho(\l_{n+1})\frac{1-\e^{-s\l_{n+1}}}{s}.
\eeqa
Laplace transforming with respect to $\l_{n+1}$ then gives
\beqa\label{eq:applapII+}
\fl \hat f_{\tau_1,\ldots,\tau_{N+1},N_t}(s;\l_1,\ldots,\l_n,u,n)
=\r(\l_1)\ldots\r(\l_n)\,\e^{-s\sum_i\l_i}
\frac{\hat\r(u)-\hat\r(u+s)}{s}.
\eeqa

\subsubsection*{{\rm (III)}}

Finally, for the third sequence,
\beqa 
\fl f_{\tau_1,\ldots,\tau_N,N_t}(t;\l_1,\ldots,\l_n,n)=
\langle\delta(\tau_1-\l_1)\ldots\delta(\tau_{n}-\l_{n})
I(t_n<t<t_{n+1})\rangle,
\eeqa
yielding
\beqa\label{eq:fIII}
\fl f_{\tau_1,\ldots,\tau_N,N_t}(t;\l_1,\ldots,\l_n,n)
=\r(\l_1)\ldots\r(\l_n)\,\int_0^\infty{\rm d}a\,p_0(a)\,\delta\left(\sum_{i=1}^n\l_i+a-t\right),
\eeqa
which can alternatively be obtained from~(\ref{eq:fI}) or~(\ref{eq:fII}).
In Laplace space
\beq
\fl\lap{t} f_{\tau_1,\ldots,\tau_N,N_t}(t;\l_1,\ldots,\l_n,n)
=\r(\l_1)\ldots\r(\l_n)\,\e^{-s\sum_i\l_i}\frac{1-\hat\rho(s)}{s},
\eeq
which can consistently be derived from~(\ref{eq:applapI+}) and (\ref{eq:applapII+}), setting $u=0$ in these expressions.

Eqs.~(\ref{eq:fI}), (\ref{eq:fII}) and (\ref{eq:fIII}) are the building blocks for the analysis performed in the bulk of the text.
For short, we denote the joint probability densities associated to the different sequences 
$\cal C^\w$ by $f^\w(\dots)$:
\beqa\label{eq:bref}
\fI(t;\l_1,\ldots,\l_n,a,n)&=&f_{\tau_1,\ldots,\tau_N,A_t,N_t}(t;\l_1,\ldots,\l_n,a,n),
\nonumber\\
\fII(t;\l_1,\ldots,\l_{n+1},n)&=&f_{\tau_1,\ldots,\tau_{N+1},N_t}(t;\l_1,\ldots,\l_{n+1},n),
\nonumber\\
\fIII(t;\l_1,\ldots,\l_n,n)&=&f_{\tau_1,\ldots,\tau_N,N_t}(t;\l_1,\ldots,\l_n,n).
\eeqa

\end{document}